\documentclass{aa}
\input psfig.tex
\usepackage{graphicx}
\usepackage{natbib}

\usepackage{array}
\usepackage{graphics}
\usepackage{latexsym}
\usepackage{amssymb}
\usepackage{amsmath}
\usepackage{fancyhdr}
\bibpunct{(}{)}{;}{a}{}{,}

\begin{document}
\title{Planar distribution of the galaxies in the Local Group: \\
                    a statistical and dynamical analysis}

\author{Stefano Pasetto \&  Cesare Chiosi}
\institute{Department of Astronomy, Padova University, Vicolo
dell'Osservatorio 2, I-35122, Padova, Italy\\
\email{{pasetto\char64pd.astro.it},{chiosi\char64pd.astro.it}}}
\date{Received: September  2006;  Revised: November 2006; Accepted...}

\titlerunning{Planar Distribution of Galaxies in the Local Group}

\abstract
{

\textsc{Aims}. Basing on the projected distribution of the galaxies
in the Local Group, Sawa \& Fujimoto found that they all seem to
distribute on a rather thin plane containing the two mayor local
galaxies of the Local Group, Milky Way  and Andromeda, and all their
dwarf satellites. As their conclusion could be severely biased by
projectional distortion effects, we re-analyse the whole issue using
a different approach. In brief, adopting known data on positions and
distances, we make use of the analytical geometry and look for the
plane that minimizes the distances of all galaxies to it. A planar
distribution is indeed found that, however, does not coincide with
the plane found by Sawa \& Fujimoto. Why? The second part of this
study is devoted to answer this question and to find a dynamical
justification for the planar distribution.

\textsc{Methods}. To this aim, we apply the Hamilton Method (Minimum
Action) to investigate the dynamics of the two major system of the
Local Group, Milky Way and Andromeda,
 under the action of external forces
exerted by  nearby galaxies or groups external to the Local Group.

\textsc{Results}. We find that the planar distribution is
fully compatible with the minimum action and that  the external
force field is likely parallel to the plane. It pulls the galaxies
of the Local Group  without altering their planar distribution.
Special care is paid to evaluate the robustness of this result.

\textsc{Conclusions}. In this paper we have examined the spatial
distribution of galaxies in the Local Group. They are confined to a
plane that can be statistically and dynamically understood as the
result of the Minimum Action. The planar distribution seems to be
stable for a large fraction of the Hubble time. The external
force field, that has likely been constant over the same time
interval, does not alter the planar distribution as it is nearly
parallel to it. Effects due to undetected halos of sole Dark Matter
are briefly discussed. They could be a point of uncertainty of the
present study.
 \keywords{Local Group,  Milky Way, Andromeda, dwarf
galaxies, spatial distribution dynamics, proper motions, Minimum
Action} }

 \maketitle
\section{Introduction}\label{introduction}

Recent reviews of the state-of-art of our understanding of
the structure, past history, galaxy content, and stellar populations
in individual galaxies of the  Local Group (LG)  are by
\citet{1999A&ARv...9..273V}, \citet{1998ARA&A..36..435M} and
\citet{2001ASPC..239..280G}. However, our present knowledge
of the proper motions of Andromeda (M31) and most galaxies  of the
LG is dramatically insufficient and inadequate to correctly
reconstruct the past and present structure of the LG. As a matter of
facts, no orbital parameters can be derived with adequate precision
to constrain the space of velocities for all known objects. As a
consequence of it, the dynamical evolution of the LG galaxies, and
the dynamical origin of the  dwarf satellites  orbiting around and
interacting with the two major galaxies, i.e. Milky Way (MW) and
M31, are not clear.

Over the years, many authors addressed these issues both
observationally and theoretically,  with the aim of understanding
the evolution of the LG galaxies. Since the first observations of
\citet{1917AJ.....30..175B},  perhaps the most important steps
toward understanding the nature of the LG  is  the determination of
the proper motion and tangential motion of M31, which unfortunately
are still  poorly known. The first important study dealing  with the
evolution of the  LG, \textsl{supposedly in isolation},  is by
\citet{1959ApJ...130..705K}. The authors based their study on the
spherical gravitational collapse and expansion approximation. As
consequence of it, with the aid of the third Kepler's  law they
predicted only radial orbits for the MW and M31 from the  origins to
the present time. The possibility of a tangential motion for M31
with respect to the MW was automatically ruled out. Another model of
the LG based on pure radial orbits for M31 and MW has been
presented by \citet{1977MNRAS.181...37L}.  The authors made use of
the second law of Newton applied to the motion of the two major
components of the LG. \citet{1977ApJ...217..903Y} focused on the
determination of the circular velocity of our own galaxy using the
LG as reference system. They studied the origin of the LG angular
momentum and for the first time emphasised how the tidal force
exerted by external proto-groups could be the cause of the origin of
the LG angular momentum, and how this interpretation should be
preferred to the cosmological origin proposed by
\citet{1951pca..conf..195H}, \citet{1969ApJ...155..393P} and
\citet{1977ApJ...216..194T}. They supposed that the spin-orbit
coupling in the first stages of the Universe could occur, even if
the angular moment gained by the spin of  Dark Matter halos and then
transferred to the orbits could be too small even for the largest
eccentricity orbits \citep{1973ApJ...186..467O,
1974ApJ...193L...1O}. As these conclusions were not in
agreement with previous studies,  they spurred the first attempts to
measure the proper motions of M31 (different from zero).

\citet{1978ApJ...223..426G} first presented the equations governing
the motion of the LG members. This study,  however, still made use
the radial approximation for the relative motion of  MW and M31. In
their picture, thanks to  the Thompson scattering before the
recombination time ($z\approx1500$), the matter is coupled with
radiation and it does not present random motions but only the motion
of the Hubble flow. After recombination, the matter decouples from
radiation: therefore galaxies and  galaxy clusters can freely form
thanks to the small unstable fluctuations of the density
\citep{1974ApJ...189L..51P, 1972ApJ...176....1G}. In this context,
the equations for the motion of MW and M31 have been written for the
first time. This  yielded   an  upper limit for the redshift at
which   the proper motions originated, $z<1500$. At earlier epochs,
speaking of proper motions was not physically meaningful. These
conclusions are, however, based on the fact that the vorticity of
the CMB  as well as that  of the initial conditions are null.
Nowadays minor anisotropies of the CMB have been discovered by WMAP
\citep{2003ApJS..148....1B,
 2006ApJ...645L..89S} thus somewhat weakening the above assumptions.
 Notwithstanding the new
observational and theoretical developments, the study of
\citet{1978ApJ...223..426G} is still  a landmark  on the origin of
the LG angular momentum.

Subsequently, \citet{1981Obs...101..111L} presented the   equations
for a \textsl{non} purely radial motion of M31.  The history of the
orbital equations for  the LG galaxies became even more complicated
with  results intermediate to those by  \citet{1981Obs...101..111L}
and \citet{1982MNRAS.199...67E}.

An independent treatment of the equations of motion for the LG
members was  presented by \citet{1985MNRAS.212..163M}.
Using the solutions of the restricted 3-body problem and assuming
 the ellipticity of $\epsilon\approx0.9$ estimated by
 \citet{1982MNRAS.199...67E} and \citet{1977ApJ...217..903Y}, it was
 argued that M31 and MW are on a linear orbit and that the minor
 galaxies maintain  a planar motion throughout at all times.

\begin{figure}
\resizebox{\hsize}{!}{\includegraphics{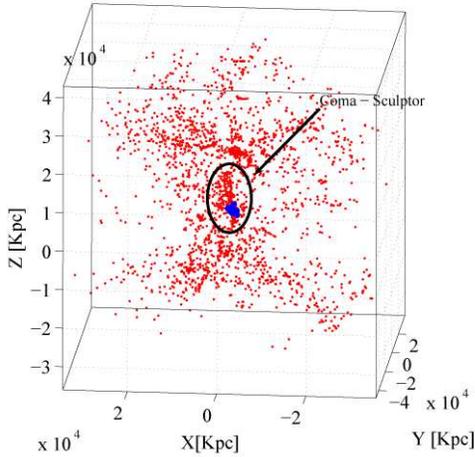}} \caption{A
sketch of the spatial distribution of the NBG galaxies in the Local
Universe.  The Sculptor-Coma cluster is highlighted and bigger dots
are used for M31, MW and the LG galaxies of
\ref{tab:LocalGroupSetOfGalaxies}. From \citet{1988ang..book.....T}}
\label{NBG01}
\end{figure}

A milestone along the road  of the evolution and formation of the LG
was set by  the Nearby Galaxies Catalog (hereinafter NBG) of
\citet{1988ngc..book.....T}, a sample of 2367 galaxies with
systematic velocities less than 3000 km/s. The spatial
 location of the NBG galaxies is shown in Fig.\ref{NBG01} to give an idea of
 structure of the Local Universe. A first systematic
investigation of the effect of the \textsl{external} galaxies on the
LG  has been  presented in \citet{1989MNRAS.240..195R}. Studying the
influence of the gravitational quadrupole interaction by external
galaxies on the LG, the authors proved that the LG is \textsl{not}
tidally isolated and  presented a new set of equations for the
motion of its galaxies. Even if the equations  and  the
proper motion of M31 given by \citet{1989MNRAS.240..195R} are not
fully correct \citep[e.g. see][]{1993MNRAS.264..865D},  the
fundamental result still remains:
 the \textsl{LG is not isolated} and the \textsl{proper motion of M31 is not
 null}.

In the same years, a flourish  of theoretical studies began
\citep{1989ApJ...344L..53P,1994ApJ...429...43P,1995ApJ...449...52P}.
In \citet{1989ApJ...344L..53P}   the equations of motion were
revised with the aid  of the Hamiltonian Principle and a new method
for tracing back the orbits of the LG galaxies  was presented and
subsequently adapted to the LG problem by
\citet{1989ApJ...345..108P} and \citet{1994ApJ...429...43P},
together with  new determinations of the tangential motion of M31
\citep{1990ApJ...362....1P,1995ApJ...449...52P}. In the series of
studies by  Peebles and collaborators, and other independent groups
as well, an interesting result was obtained
\citep{1995ApJ...454...15S,1995ApJ...449...52P,
1999ASPC..176..280S,2001ApJ...554..104P, 2001ApJ...550...87G}: the
solutions for the local minimum or stationary points of the action
are compatible with a tangential motion of M31 as large as the
radial motion.

Following  the same methodology, \citet{1993MNRAS.264..865D}
independently derived the value for the proper motion of M31
$\left(\mu_{l},\mu_{b}\right) = \left(0.126,0.0048\right)$ mas/yr.
Despite this great improvement in the knowledge of the LG, new
problems arose regarding its cosmological origins as noted  by
\citet{1997NewA....2...91G}. They pointed out the strong dependence
of the theoretical models on the cosmological parameters for the
peculiar velocities of galaxies  in  CDM context. The same
dependence has been suspected for some observational constraints,
e.g. for the distances or the radial velocities
\citep{1993AJ....105..886V, 1994AJ....107.2055B}. Recently a new
determination of the mass of M31 has been derived by
\citet{2000MNRAS.316..929E}. In contrast with all previous dynamical
estimates, the mass assigned to M31 is smaller than the mass of MW.
This result can be  explained recalling that  it stems from the
notion that the velocity vector of M31 is purely radial and  that
the mass is derived from  velocity arguments. One may argue that the
lower mass assigned to M31 could be taken as an hint that the motion
of M31 is not merely radial.

Moreover, attempts were made to find the spatial distribution of the
LG galaxies, e.g.  \citet{2000AJ....119.2248H} suggested that the
galaxies crowd on a  highly flattened ellipsoid with axial ratios
$\left(a,b,c\right) = \left(1.00, 0.51, 0.19\right)$ which to a good
approximation can be considered as a  planar distribution. Another
problem under investigation was  the anisotropic distribution of
inner sub-haloes with respect to larger haloes  in relation to the
Holmberg effect \citep{2004MNRAS.348.1236S}. General consensus on
this issue has not yet been reached \citep{2004ApJ...603....7K}. In
particular it is unclear and matter of debate whether
disruption-tidal effects can create the apparent polar alignment of
the dwarf satellites around the mother galaxy or, for the particular
case of the LG, the  position of the dwarf galaxies is the
consequence of  peculiar directions of  pre-existing cosmological
filaments. See for instance Fig.\ref{NBG01} which displays the
spatial distribution of the NBG galaxies. In a recent paper,
\citet{2005A&A...431..517K} suggest a planar distribution of the
satellites of MW, which however could also be explained as a
consequence of the distribution of sub-haloes
\citep{2005ApJ...629..219Z} in the early cosmological stages
\citep{2005A&A...437..383K}. Another important result for
the sub-halo alignment in cosmological simulations was obtained by
\citet{2005MNRAS.363..146L}. The same problem is under
investigation for M31 \citep{2006AJ....131.1405K}. Finally,
\citet{2005PASJ...57..429S} have recently proposed a new dynamical
model based on the existence  an orbital plane for M31 and
MW.

In this paper we address some of the above issues. The plan is as
follows. In Sect. \ref{plane} we examine the spatial distribution of
the LG galaxies: starting from the analysis of
\citet{2005PASJ...57..429S} and using a different procedure
(Principal Component Analysis) we find that a common plane really
exists but different from the one found by
\citet{2005PASJ...57..429S}. We then proceed to  study the dynamics
of the LG. In Sect. \ref{minimization} we present the Hamilton
method of Constrained Minimum Action.  In Sect. \ref{catalogs} we
present the samples of galaxies in the Local Universe we have
adopted to study the dynamics of the LG and the effects on this by
nearby galaxies. In Sect. \ref{results} we minimize the Action and
present the results. In Sect. \ref{Robustness} we examine in detail
the robustness of the procedure  we have adopted.  In Sect. \ref{EF}
we evaluate the effect of external galaxies on the dynamics of the
LG. The main result is that the orbits of the dominant galaxies of
the LG  are dynamically consistent with a plane that within the
uncertainties coincides with the one derived in Sect. \ref{plane}
from mere geometrical considerations. This is indeed the plane that
minimizes the action. Several important implications of this finding
are discussed in the above sections. Finally, in Sect. \ref{concl}
we draw some concluding remarks.

\section{Is there a plane on which LG galaxies crowd?}\label{plane}

As already mentioned,  from the visual inspection of the
sky-projected spatial distribution of the LG galaxies
\citet{2005PASJ...57..429S} suggest the existence of an orbital
plane for MW, M31 and satellites. However the idea is not new. 2-D
studies have  already been employed to establish planes in the LG.
Limiting ourselves to a few examples, \citet{2000AJ....119.2248H}
suggested a flat  ellipsoid which is not too different from a plane;
studies of the LG dynamics, see for instance
\citet{1959ApJ...130..705K} and \citet{1989MNRAS.240..195R}. Finally
\citet{1979ApJ...228..718K}, \citet{ 1999IAUS..192..447G} and
\citet{1995AJ....110.1664F} suggested that the satellite dwarf
galaxies of the MW and M31 lie on planes.

We suspect, however,  that the plane found by
\citet{2005PASJ...57..429S} largely results from distortion effects
affecting the visual inspection of spatial structures. Therefore, in
this section we reconsider the whole subject and check whether such
a plane can be derived from other objective methods.

Our analysis of the galaxies distribution stands on the
\citet{1998ARA&A..36..435M} sample of LG galaxies (see the entries
of Table \ref{tab:LocalGroupSetOfGalaxies}). Whenever required the
radial velocities are taken from \citet{1988ang..book.....T},
however updated to more recent determinations if available (see
below). The same sample has been used by
 \citet{2005PASJ...57..429S}.

\begin{figure}
\resizebox{\hsize}{!}{\includegraphics{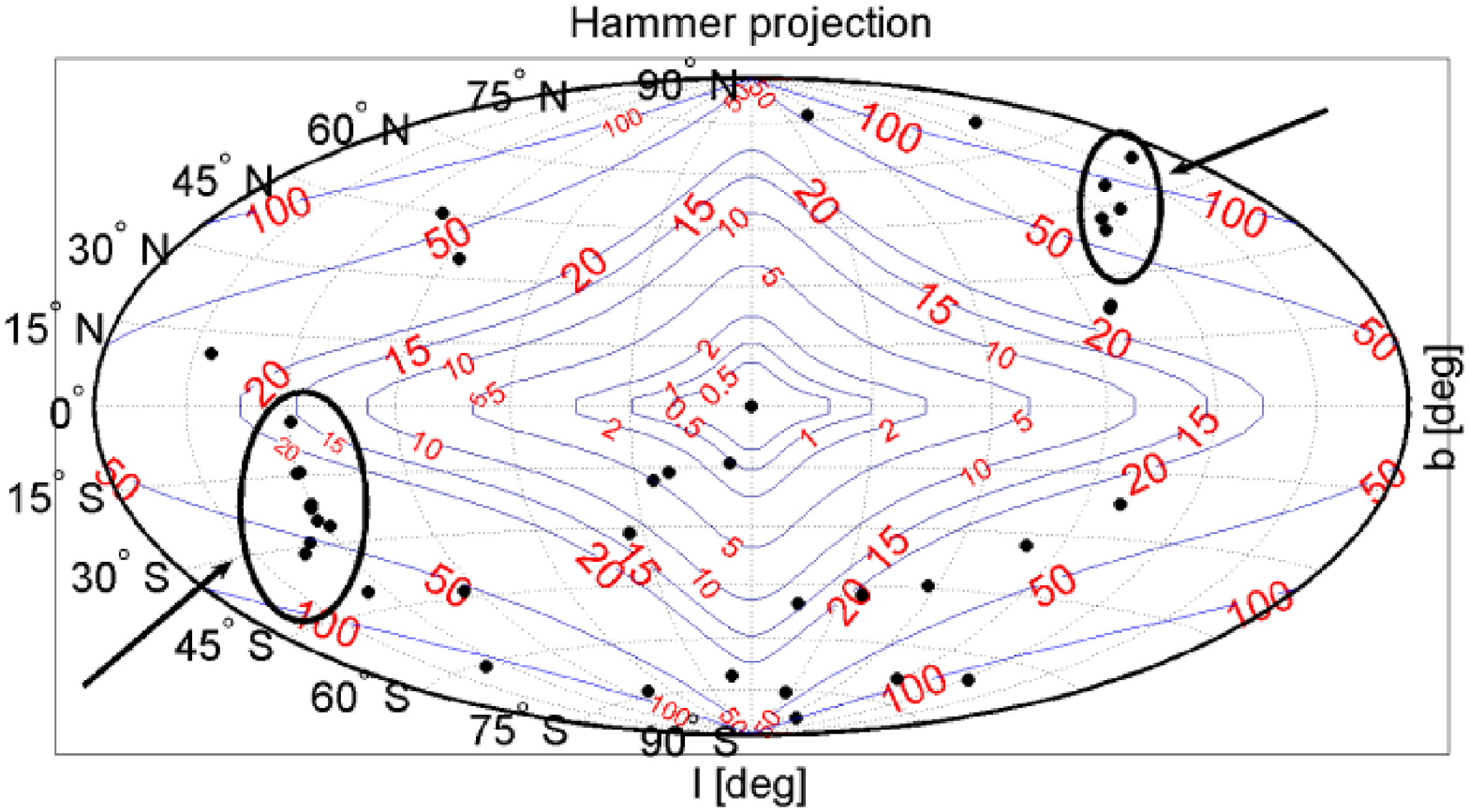}}
\caption{Hammer's projection with the amplitude of distortion (in
percent) according to the code 1=no distortion, 100=extremely high
distortion. The heavy dots are the galaxies of Table
\ref{tab:LocalGroupSetOfGalaxies}.  The arrows indicate the regions
of larger concentration of objects that could be interpreted  as
hints of  underlying 3-D structures. They, however, occur in regions
of large distortion} \label{PHammer03}
\end{figure}

\begin{figure}
\resizebox{\hsize}{!}{\includegraphics{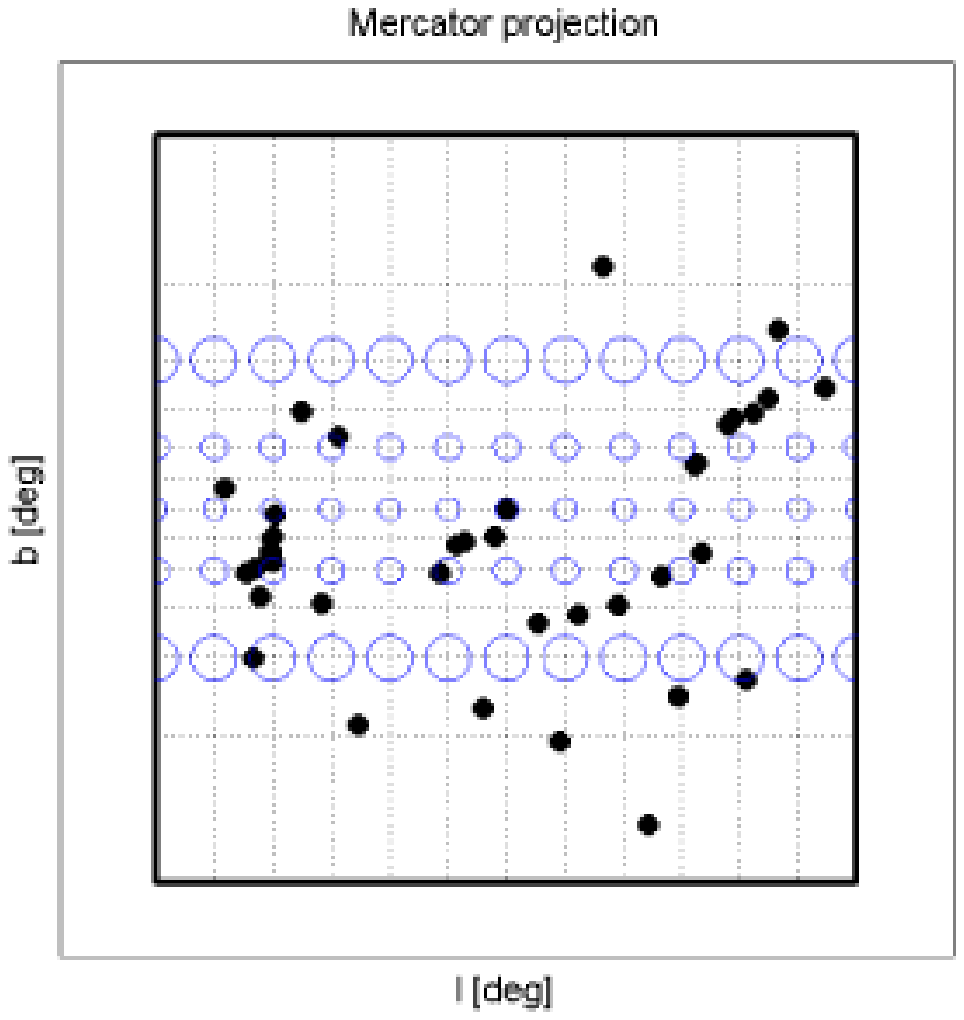}}
\caption{Mercator's projection with Tissot's circles. The Mercator
projection provides a good representation near the equatorial plane
and a very poor one towards the poles. The lack of galaxies near the
galactic plane is due to Galactic absorption. Clearly the apparent
distribution is not uniform. There seem to be several 2-D structures
that could perhaps hint for real 3-D structures. However, the same
remark made on Fig.\ref{PHammer03} would apply also here. }
\label{Mercator}
\end{figure}

\begin{figure}
\resizebox{\hsize}{!}{\includegraphics{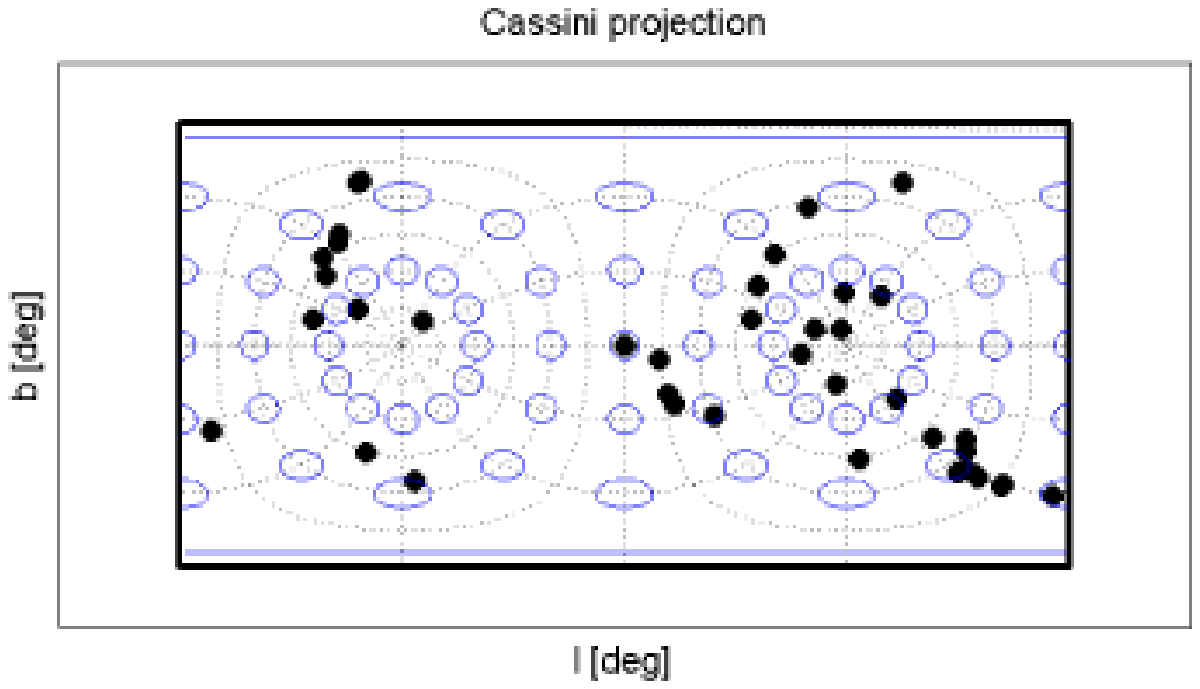}}
\caption{Cassini's projection with Tissot's circles. The North and
South Galactic Poles are  now well  represented. Once again a few
2-D structures can be noticed. They, however are located in regions
of high distortion so that the same remark made on
Fig.\ref{PHammer03} would apply also here} \label{Cassini}
\end{figure}

 Following
\citet{2005PASJ...57..429S} we start by looking at the projected
distribution of the LG members on the celestial sphere.  It is soon
evident that this way of proceeding is severely biased by the type
of projection in use. To illustrate the point, in
Fig.\ref{PHammer03} we show the Hammer projection on which the
contours of projection-induced distortion on the map are superposed.
The distortion is measured by the ratio of the infinitesimal areas
given by the jacobian of the transformation matrix between the
cartesian coordinate system and the projected coordinate system. The
amplitude of distortion is given in percent, i.e. normalised to an
un-distorted projection. As a matter of fact the major clustering of
objects that could suggest peculiar spatial structures such as
filaments, planes etc, actually occurs in regions of the Hammer
plane with distortion amounting to about 50 - 75 percent. Therefore,
visual inspections of the projected distribution to infer spatial
structures could  lead to wrong conclusions. Other kinds of
projection do not make a better job. For instance in
Fig.\ref{Mercator} and Fig.\ref{Cassini} we show the Mercator and
Cassini projections, respectively.  We have added the Tissot's
circles, which are equal circles on the celestial sphere to better
visualise the effects of projectional distortions. In principle, the
combination of these two projections could improve upon the correct
view of different areas on the celestial sphere. In the Mercator
projection all the lines with the same angles with respect to the
meridians are straight lines, and thus  reduce the distortion near
the equatorial plane of the MW ($l < 86^\circ$ is conventionally
adopted). Unfortunately, as all dwarf galaxies of the LG have $|b| <
84^\circ$,  in the direction $\left(l,b\right) \approx
\left(0,0\right)$ they are masked by the Galactic Plane,  even the
Mercator projection does not help us very much. The Cassini
projection ensures instead a good approximation toward the Galactic
Pole.  Once again the few apparent structures that could hint for
real 3-D structures are located near regions of large distortion.
Therefore,  the main result of this first analysis is  that the 3-D
structure of the LG when projected onto 2-D surfaces could mimic
planar distributions that may not exist in reality.

\begin{table*}
\caption{Parameters of LG galaxies used in this study. }
\centering
\begin{tabular}{|c|c|c|c|c|c|c|c|c|}
\hline
Galaxy Name &Other name&l& b& r& Group&x&y&z \\
\hline
WLM        &DDO 221     &75.9  &-73.6 &925    &LGC     &-55.1  &253.3   &-887.4  \\
NGC55      &            &332.7 &-75.7 &1480   &LGC     &-316.3 &-167.7  &-1434.1 \\
IC10       &UGC192      &119   &-3.3  &825    &M31     &407.8  &720.4   &-47.5   \\
NGC147     &DDO3        &119.8 &-14.3 &725    &M31     &357.6  &609.6   &-179.1  \\
And III    &            &119.3 &-26.2 &760    &M31     &342.2  &594.7   &-335.5  \\
NGC185     &UGC396      &120.8 &-14.5 &620    &M31     &315.9  &515.6   &-155.2  \\
NGC205     &M110        &120.7 &-21.1 &815    &M31     &396.7  &653.8   &-293.4  \\
M32        &NGC221      &121.2 &-22   &805    &M31     &395.1  &638.4   &-301.6  \\
M31        &NGC224      &121.2 &-21.6 &770    &M31     &379.4  &612.4   &-283.5  \\
And I      &            &121.7 &-24.9 &805    &M31     &392.2  &621.2   &-338.9  \\
SMC        &NGC292      &302.8 &-44.3 &58     &MW      &-14.0  &-34.9   &-40.5   \\
Sculptor   &            &287.5 &-83.2 &79     &MW      &5.7    &-8.9    &-78.4   \\
LGS3       &Pisces      &126.8 &-40.9 &810    &M31     &375.2  &490.2   &-530.3  \\
IC1613     &DDO8        &129.8 &-60.6 &700    &M31/LGC &228.5  &264.0   &-609.8  \\
And II     &            &128.9 &-29.2 &525    &M31     &296.3  &356.7   &-256.1  \\
M33        &NGC598      &133.6 &-31.3 &840    &M31     &503.5  &519.8   &-436.4  \\
Phoenix    &            &272.2 &-68.9 &445    &MW/LGC  &2.4    &-160.1  &-415.2  \\
Fornax     &            &237.1 &-65.7 &138    &MW      &39.3   &-47.7   &-125.8  \\
EGB0427+63 &UGCA92      &144.7 &10.5  &1300   &M31     &1051.7 &738.6   &236.9   \\
LMC        &            &280.5 &-32.9 &49     &MW      &1.0    &-40.5   &-26.6   \\
Carina     &            &260.1 &-22.2 &101    &MW      &24.6   &-92.1   &-38.2   \\
Leo A      &DDO69       &196.9 &52.4  &690    &MW/N3109&411.3  &-122.4  &546.7   \\
Sextans B  &DDO70       &233.2 &43.8  &1345   &N3109   &590.0  &-777.3  &930.9   \\
NGC3109    &DDO236      &262.1 &23.1  &1250   &N3109   &166.5  &-1138.9 &490.4   \\
Antlia     &            &263.1 &22.3  &1235   &N3109   &145.8  &-1134.4 &468.6   \\
Leo I      &            &226   &49.1  &250    &MW      &122.2  &-117.7  &189.0   \\
Sextans A  &DDO75       &246.2 &39.9  &1440   &N3109   &454.3  &-1010.8 &923.7   \\
Sextans    &            &243.5 &42.3  &86     &MW      &36.9   &-56.9   &57.9    \\
Leo II     &DDO93       &220.2 &67.2  &205    &MW      &69.2   &-51.3   &189.0   \\
GR8        &DDO155      &310.7 &77    &1590   &GR8     &-224.7 &-271.2  &1549.2  \\
Ursa Minor &DDO199      &105   &44.8  &66     &MW      &20.6   &45.2    &46.5    \\
Draco      &DDO208      &86.4  &34.7  &82     &MW      &4.3    &67.3    &46.7    \\
Sagittarius&            &5.6   &-14.1 &24     &MW      &-14.7  &2.3     &-5.8    \\
SagDIG     &UKS1927-177 &21.1  &-16.3 &1060   &LGC     &-940.7 &366.3   &-297.5  \\
NGC6822    &DDO209      &25.3  &-18.4 &490    &LGC     &-411.9 &198.7   &-154.7  \\
DDO 210    &Aquanus     &34    &-31.3 &800    &LGC     &-558.2 &382.2   &-415.6  \\
IC5152     &            &343.9 &-50.2 &1590   &LGC     &-969.4 &-282.2  &-1221.6 \\
Tucana     &            &322.9 &-47.4 &880    &LGC     &-466.6 &-359.3  &-647.8  \\
UKS2323-326&UGCA 438    &11.9  &-70.9 &1320   &LGC     &-414.1 &89.1    &-1247.3 \\
Pegasus    &DDO216      &94.8  &-43.5 &955    &LGC     &66.5   &690.3   &-657.4  \\
        \hline
        \end{tabular}
\begin{minipage}{0.9 \textwidth}
\footnotesize{ The coordinate system is centered on the MW.  The X
axis points from the centre of MW to the Sun, located at
$R_\odot=8.5\,\,kpc$, the Z axis points to the North Galactic Pole,
and the third axis  forms a right handed coordinate system. Distances are in kpc. The
same sample has been used by \citet{2005PASJ...57..429S}. We do not
show  the radial velocities because they are not used in this paper.
} \\
\end{minipage}
\label{tab:LocalGroupSetOfGalaxies}
\end{table*}

For the above reasons, we  decided to approach  the problem in a
different way, i.e. making use of  the analytical geometry to
investigate the idea of \citet{2005PASJ...57..429S}. We start
\textsl{by assuming} that the  \citet{2005PASJ...57..429S}
conclusion is true, i.e. LG galaxies crowd on a \textsl{plane}, but
check it. The method we have adopted is ultimately the Principal
Components Analysis \citep[see for instance][]{PCA}. Given
\textsl{N} galaxies (of the LG), with coordinates ${\bf{x}}^{\left(
g \right)}, g = 1...N$ we want to find the plane that  best
approximates  their spatial distribution. In other words, given
\textsl{N} points, which are the principal directions of the plane
best approximating  these points?

The Hessian normal representation of a plane is  suited to our aims

\begin{equation}
    {\bf{\hat n}} \cdot {\bf{x}} =  - c  \, ,
\end{equation}

\noindent where the 3-D vector $\bf{\hat n}$ is the unitary vector
orthogonal   to the plane, and \textsl{c}  the distance from the
origin of the coordinate system in use. The reason for choosing this
geometrical representation of the plane instead of the classical one
$ax + by + cz + d = 0$ is that in the Hessian formalism the distance
of any given point to the plane ${\bf{\tilde x}}$ is particularly
simple:

\begin{equation}
    D_i = {\bf{\hat n}} \cdot {\bf{\tilde x}}_i  + c \, .
\label{distance}
\end{equation}

\noindent Then we   minimize the equation for the distance $D$ for
all points

\begin{equation}
\label{distancef} D = \sum\limits_{g = 1}^N {\left| {{\bf{n}} \cdot
{\bf{x}}^{\left( g \right)}  + c}    \right|^2 } \, .
\end{equation}

\noindent The simplest way  is given by the Lagrange multipliers.
First  we introduce the condition ensuring the ortho-normality of
the vector $\bf{\hat n}$

\begin{equation}
\label{constr1}
    \left| {\bf{n}} \right|^2  - 1 = 0
\end{equation}

\noindent and get  the system of equations

\begin{equation}
\label{sys1}
\left\{ \begin{array}{l}
 \begin{array}{*{20}c}
   {\sum\limits_{g = 1}^N {\left[ {\left( {{\bf{n}} \cdot {\bf{x}}^{\left( g \right)}  + c}
    \right)x_i^{\left( g \right)} } \right]}  = \lambda n_i } & {\forall i = 1...3}  \\
\end{array} \\
 \sum\limits_{g = 1}^N {\left( {{\bf{n}} \cdot {\bf{x}}^{\left( g \right)}  + c} \right)}  = 0 \\
 \left| {\bf{n}} \right|^2  - 1 = 0 \\
 \end{array} \right.
\end{equation}

\noindent  \citep[see e.g.][]{DeMarco}. It follows that distances
are linked to the eigen-values of the eigen-system $ {\bf{A}} \cdot
{\bf{n}} = \lambda {\bf{n}}$ that can be derived from the system
(\ref{sys1}), where the matrix $\bf{A}$ is given by

\begin{equation}
{\bf{A}}_{ij}  \equiv \sum\limits_{g = 1}^N {\left(
{\left({x_j^{\left( g \right)} - \frac{1}{N}\sum\limits_{h = 1}^N
{x_j^{\left( h \right)} } } \right)x_i^{\left( g \right)} } \right)}
\, .
\end{equation}

\noindent The resulting eigen-values, corresponding to 3 values for
the minimum of $D$, are

\[
\begin{array}{l}
 D^{\left( 1 \right)}  = \lambda ^{\left( 1 \right)}  = {\rm{2}}
 {\rm{.01}}{\bf{ \times }}10^7  \\
 D^{\left( 2 \right)}  = \lambda ^{\left( 2 \right)}  = {\rm{1}}
 {\rm{.05}}{\bf{ \times }}10^7  \\
 D^{\left( 3 \right)}  = \lambda ^{\left( 3 \right)}  = {\rm{3}}
 {\rm{.88}}{\bf{ \times }}10^6  \, ,\\
 \end{array}
\]

\noindent which yield the eigen-vectors:

\begin{equation}
\label{vectors1}
    \begin{array}{l}
 {\bf{n}}^{\left( 1 \right)}  = \left( {{\rm{0}}{\rm{.19}}{\rm{,  - 0}}
 {\rm{.51}}{\rm{, 0}}{\rm{.83}}} \right) \\
 {\bf{n}}^{\left( 2 \right)}  = \left( {{\rm{0}}{\rm{.66}}{\rm{, 0}}
 {\rm{.69}}{\rm{, 0}}{\rm{.27}}} \right) \\
 {\bf{n}}^{\left( 3 \right)}  = \left( {{\rm{0}}{\rm{.72}}{\rm{,  - 0}}
 {\rm{.50}}{\rm{,  - 0}}{\rm{.47}}} \right) \, .\\
 \end{array}
\end{equation}

\begin{figure*}
\resizebox{\hsize}{!}{\includegraphics{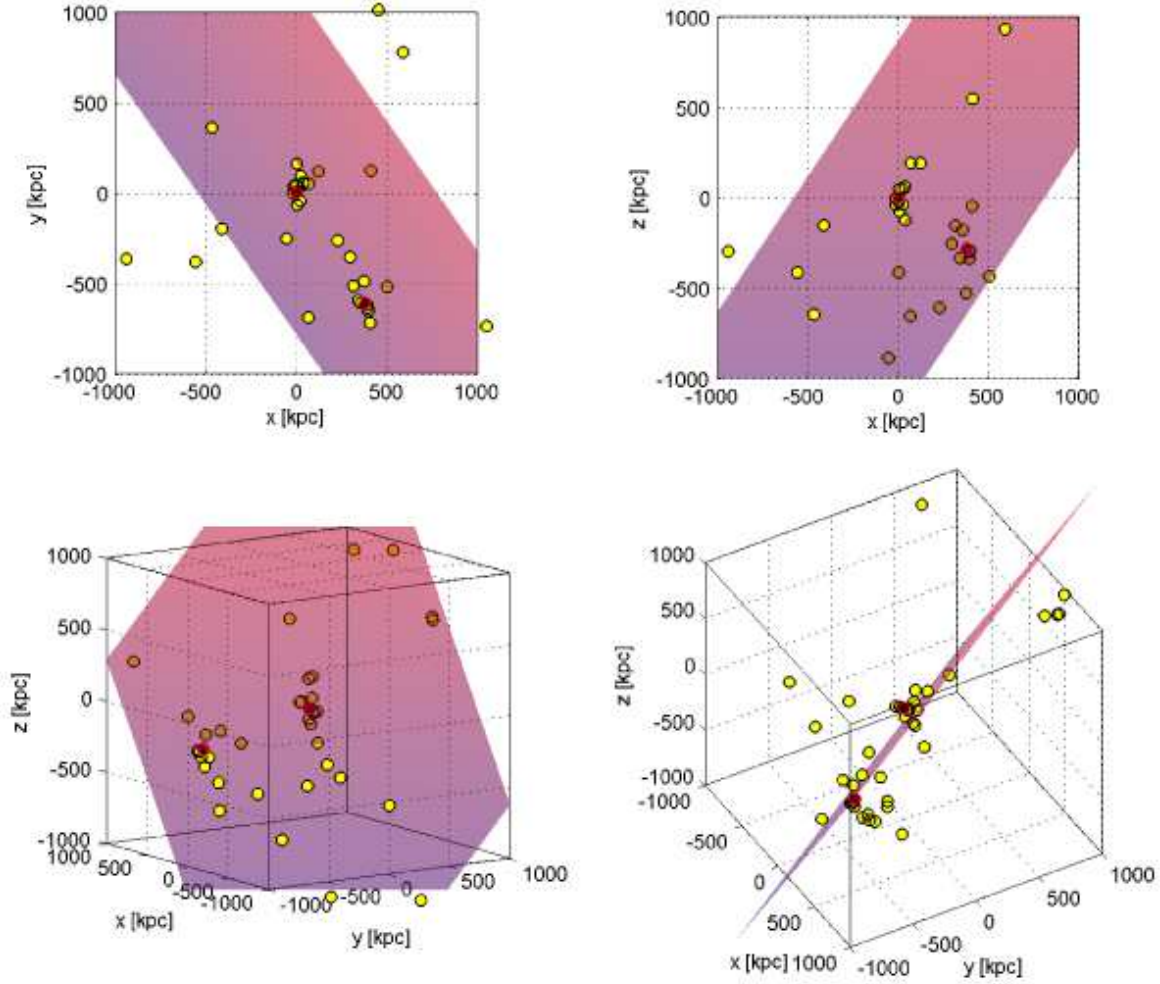}}
\caption{Projection onto the coordinate planes (X,Y) and
(X,Z) as indicated, and two 3-D views (bottom left and right panels)
of the our plane best minimizing the distance (location) of the LG
galaxies.
 The (X,Y) projection (top left panel) is
parallel to the equatorial plane of the MW. The big red filled
circles are MW and M31.    The yellow circles are the dwarf
satellites on either sides of the plane}. Looking at the 3-D view in
which the plane is seen edge-on (bottom right), MW and M31 are very
close to it and within the uncertainty can be considered as
belonging to it. See the text for details  \label{PlanesPANEL01}
\end{figure*}

Each of these vectors corresponds to a plane.  Therefore we  have 3
orthogonal planes out of which we  have to  select the one yielding
the absolute minimum distances. The planes are referred to an
orthogonal reference system co-aligned with the MW located at the
origin. Clearly the plane to choose is the third one.
Finally, for the sake of the comparison with other studies,  it is
worth noting here that the normal vector of this plane is along the
direction $(l,b)=(124^\circ, -28^\circ)$.

The situation of the plane is shown in  Fig.\ref{PlanesPANEL01}
where the top left and right panels show the LG galaxies projected
onto the plane of the MW and perpendicular to it, whereas the bottom
left and right panels show the same galaxies projected face-on and
edge-on  onto the plane minimizing $D$. From the lower right panel
of Fig.\ref{PlanesPANEL01} we see that the two dominant galaxies (MW
and M31) do not strictly belong to the plane. However, considering
the  uncertainties in the position of each galaxy, we may reasonably
conclude that \textit{also the two major galaxies of the system
belong to the plane}\footnote{A detailed \textsl{astrophysical }
explanation (\textsl{not geometrical}) for the errors in the
definition of the plane will be given later in this section.}.

Starting from  this  first result we may also assume  that the
\textsl{geometrical plane contains the barycentre  of the system}.
This in not true in general but limited to the LG. It follows that
the distribution of the baryonic (and dark) matter is symmetric with
respect to  this plane. Furthermore, the fact that the  two most
massive galaxies belong to the plane could perhaps indicate that the
effect, if any, of still undetected massive bodies influencing the
LG dynamics is small and likely was such in past. On the problem of
the mass tracers \citep{1983ApJ...267..465D, 1988ApJ...333L..45W} we
will come back in Sect. \ref{minimization}. To conclude  we may
provisionally \textsl{assume} that

\begin{enumerate}
\item No pure dark bodies should exist that could significantly affect
the dynamics of the LG (see Sect. \ref{minimization} for more
details). Therefore,  the dominant galaxies (with the
largest mass) determining the  motion of the dwarf satellites are MW
and M31. In other words, in the  CDM hierarchical  accretion
scenario, if there were a pure dark body as massive as the MW or M31
halos, we would expect the dwarf  galaxies (which owing to  their
small mass are good traces of the global gravitational potential)
not to lie on the same plane containing MW and M31.

\item The problem of the small number of known dwarf galaxies,
 \citep[see e.g.][]{1999ApJ...516..530K, 1999ApJ...524L..19M}, as
compared to the expected one is not relevant for the dynamics of the
LG, see e.g. \citet{1998ARA&A..36..435M} or
\citet{1999A&ARv...9..273V}. In other words, even in
presence of a large number of dwarf satellites, less massive that
about $10^{7}\, \, M_{\odot}$, the dynamics of the known LG dwarf
satellites is not strongly affected. There is indeed no reason for a
group of about 100 unseen satellites to cluster in a small  region,
so that they could deeply bias the plane determination. This is
particularly true if, in virtue of item 1,  only dark bodies are
involved.
\end{enumerate}

\begin{figure}
\resizebox{\hsize}{!}{\includegraphics{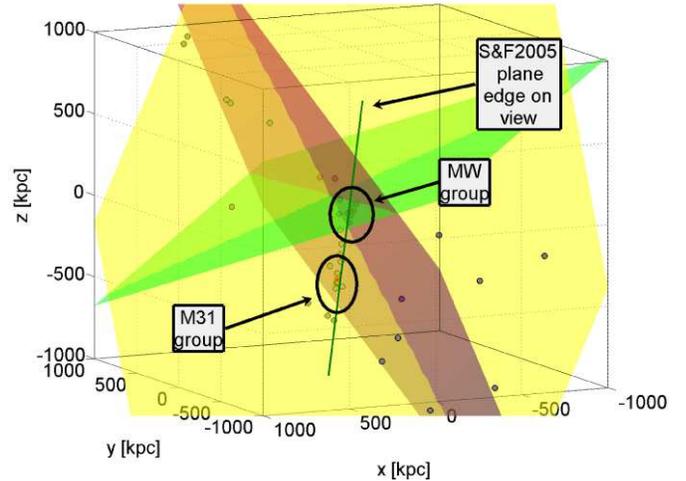}} \caption{The
3-D view of the three planes we have found with Eqn.
(\ref{vectors1}) together with the plane found by
\citet{2005PASJ...57..429S} seen edge-on and the galaxies of the LG.
The two encircled zones highlight the position of MW and M31 and
their satellites. None of our planes coincides with that of
\citet{2005PASJ...57..429S}. See the text for more details}
\label{MinPlaneconSF200504}
\end{figure}

\begin{figure}
\resizebox{\hsize}{!}{\includegraphics{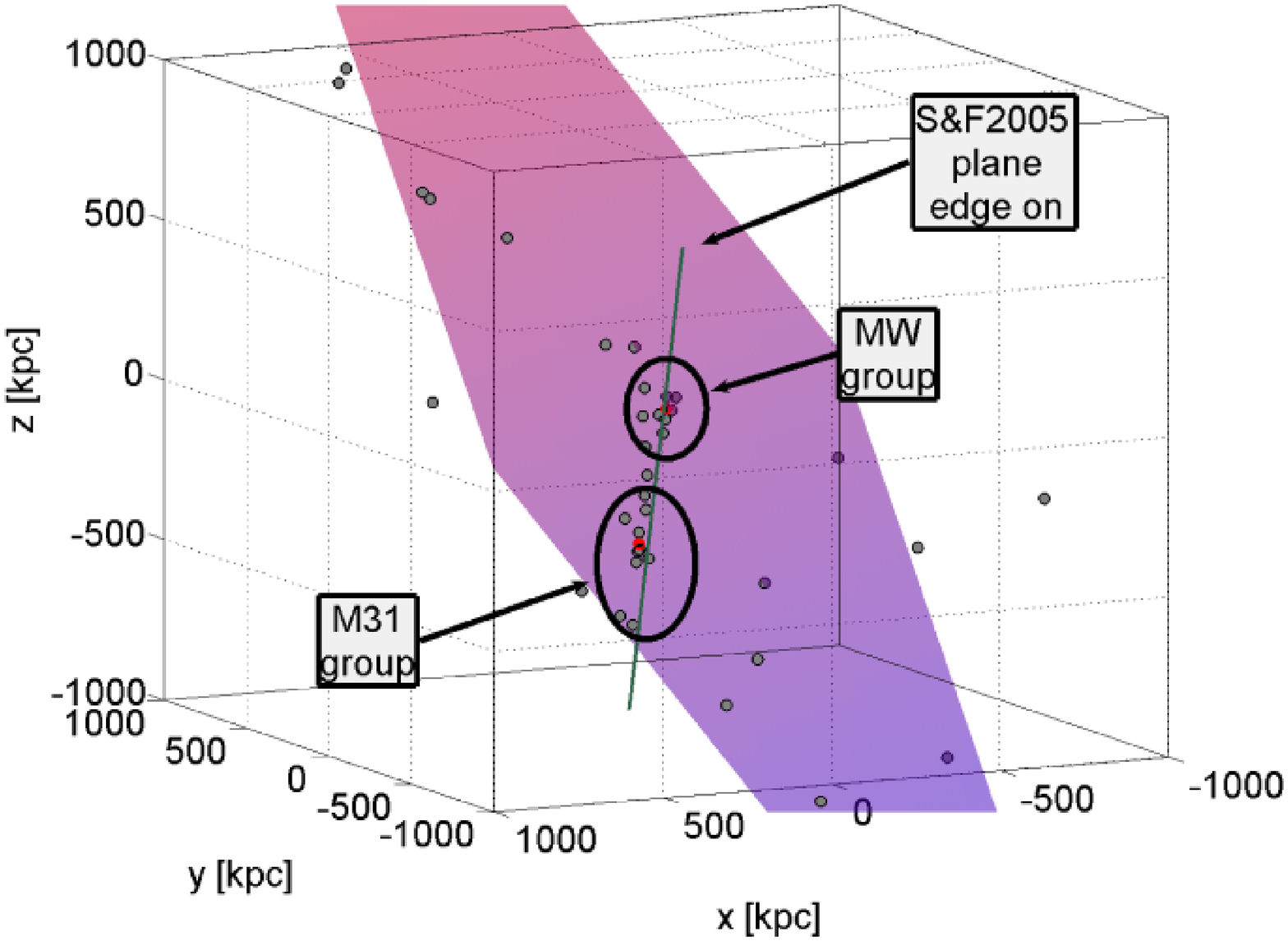}}
\caption{3-D-view of the LG as observed from the same view angle
adopted by   \citet{2005PASJ...57..429S}, i.e.
$(l,b)=(296^{\circ},-11^{\circ})$. The hatched plane is our best
plane minimizing the distances and containing also MW and M31. The
red circles are MW and M31, the yellow circles are the dwarf
galaxies above the plane, whereas the dark circles are those below
it. Once again the MW and M31 groups are indicated. The straight
line is the \citet{2005PASJ...57..429S} plane seen edge-on. See the
text for details } \label{MinM31MWPlane01}
\end{figure}

The most interesting result is however that none of the planes we
have found  coincides with that of \citet{2005PASJ...57..429S}. To
better illustrate the point,  using the same view angle adopted in
Fig.3 of \citet{2005PASJ...57..429S}, i.e.
$(l,b)=(296^{\circ},-11^{\circ})$, in Fig.\ref{MinPlaneconSF200504}
 we display the three planes we have derived from Eqn.
(\ref{vectors1}) (only one minimizes the distances of the galaxies
to it) and the plane found by \citet{2005PASJ...57..429S}. To add
more information,  in Fig.\ref{MinM31MWPlane01} we show only our
best plane and that of  \citet{2005PASJ...57..429S}.  No one of our
planes coincide with theirs. Why?

To strengthen our result we extend the above analysis as follows.
Let us suppose that all the  galaxies of the LG are located on a
plane. It goes without saying that this plane  contains also the
vector joining MW and M31. Now we may turn the argument around and,
fixing the straight line joining MW and M31, we may consider the
manyfold of planes passing through this line. Therefore our previous
analysis and result are now translated into finding the plane of the
manyfold that minimizes the  distances to it of all remaining
galaxies. We would like to note here that the degree of freedom to
our disposal (rotation of the plane around the MW-M31 line) is not a
mere mathematical problem: it is indeed due to the yet unsolved
question of the tangential motion of M31 with respect to the MW
(widely discussed in Sect. \ref{introduction}). Similar, though  not
identical reasoning was followed by \citet{2005PASJ...57..429S} when
deriving the orbits of M31 and MW by means of  the projection
analysis. Fixing the plane is equivalent to reduce the degree of
freedom for the motion of M31 and to set a constraint on the
dynamics of the LG as a whole (Pasetto \& Chiosi 2006, in
preparation).

If the MW-M31 vector  belongs to the plane in question (reasonable
assumption),  the motion is mainly  \textsl{central} and
\textsl{plane}. Small perturbations of the central motion will be
investigated in the Sect. \ref{setofgalaxies}. We are interested
here to investigate the possibility that the orbital plane of the
two main galaxies might  be  \textsl{related} to the plane we have
just derived from mere geometrical considerations. As this result is
not the rule, different determinations of the proper motion of M31
are possible, see for instance \citet{1989MNRAS.240..195R} and
\citet{2005PASJ...57..429S}, depending on the kind of  different
physical considerations that guided the authors towards the
reconstruction of  the dynamical history of the LG.

The target now is to minimize the Eqn. (\ref{distancef}) imposing
the constraint given by Eqn. (\ref{constr1}) and the additional one
 that the normal to the minimization plane has to be orthogonal to
the MW-M31 vector ${\bf{a}} = \left( {a_x ,a_y ,a_z }
\right)$. The latter constraint is expressed by the relation

\begin{equation}
    {\bf{n}} \cdot {\bf{a}} = 0 \, .
\end{equation}

\noindent All this leads to the system of equations (the
demonstration is given in Appendix A)

\begin{eqnarray}\label{sys2}
    \left\{ \begin{array}{l}
 2\sum\limits_{g = 1}^N {n_i x_i^g x_j^g }  = \lambda _1 2n_j  + \lambda _2 a_j  \\
 n_i n_i  - 1 = 0 \\
 n_i a_i  = 0  \,\, .\\
 \end{array} \right.
\end{eqnarray}

\noindent The solutions for the  normals of the plane are:

\begin{eqnarray}
\label{vectors2}
    \begin{array}{l}
 {\bf{n}}^{\left( 1 \right)}  = \left( { \pm 0.58, \pm 0.01, \pm 0.81} \right) \\
 {\bf{n}}^{\left( 2 \right)}  = \left( { \pm 0.64, \mp 0.61, \mp 0.45} \right) \\
 \end{array}
\end{eqnarray}

\noindent  that with the aid of the Eqn. (\ref{distancef}) give

\begin{eqnarray}
    \begin{array}{l}
 D^{\left( 1 \right)}  = 1.39{\bf{ \times }}10^7  \\
 D^{\left( 2 \right)}  = 3.11{\bf{ \times }}10^6  \,\, .\\
 \end{array}
\end{eqnarray}

\noindent The second solution seems to be  better than the
first one.  This plane has the normal along the direction
$(l,b)=(133^\circ,-27^\circ)$.

\begin{figure}
\resizebox{\hsize}{!}{\includegraphics{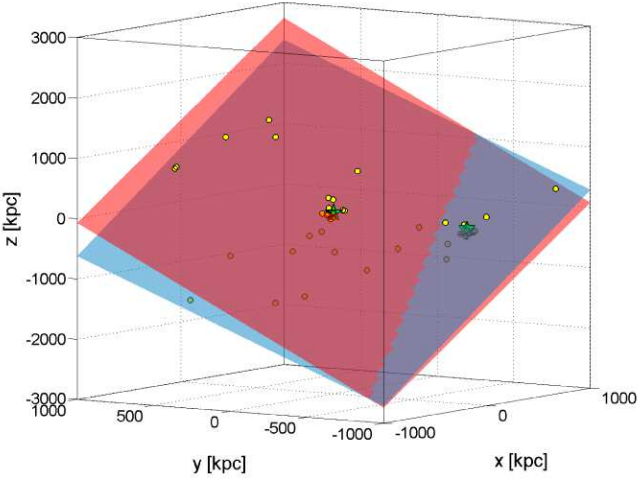}} \caption{The
two planes minimizing the spatial distribution of galaxies in the
LG. One is the best plane derived from solving the system of
equation (\ref{sys1}) and given by the eigen-vectors of Eqn.
(\ref{vectors1}) that has already been presented, the other is
obtained from the system of Eqns. (\ref{sys2}) and the eigen-vectors
of Eqn. (\ref{vectors2}). This second solution assumes that the
vector MW-M31 lays on the plane.  Within the uncertainty the two
planes coincide} \label{Min1Min2_01}
\end{figure}

Clearly the two planes defined by the  systems of Eqns.
(\ref{vectors1}) and (\ref{vectors2}) are not the same.  Indeed
system (\ref{vectors2}) is a simplification of   system
(\ref{vectors1}). However, the difference between the two planes is
small as  shown in Fig.\ref{Min1Min2_01}.  This amply justifies the
use of the results  from system (\ref{vectors2}) in the analysis
below.

\section{Dynamics of the LG from the Hamilton method}\label{minimization}

To investigate the dynamics of the LG we use the Hamilton's
principle of Minimum Action. The method was developed and adapted to
cosmological problems in Lagrangian description by
\citet{1989ApJ...344L..53P}, and its applicability  was tested by
\citet{1980lssu.book.....P} and \citet{1960QB981.B72......}. The key
ingredients are the boundary conditions applied to the initial and
end point of each trajectory.

Each mass point (galaxy) evolves under the gravitational potential
of an expanding Universe constrained by the condition

\begin{eqnarray}
\label{conditions1}
    \left\{ \begin{array}{l}
 \delta {\bf{x}}_i \left( {t = 0} \right) \ne 0 \wedge
 \mathop {\lim }\limits_{a \to 0} a^2 {\bf{\dot x}}_i  = 0 \\
 \mathop {\lim }\limits_{t \to t_0 } \delta {\bf{x}}_i  = 0 \\
 \end{array} \right.
\end{eqnarray}
\noindent and the action \citep{1989ApJ...344L..53P}
\begin{equation}
\label{action1}
\begin{array}{l}
 S = \int_0^{t_0 } {\left[ {\sum\limits_{i = 1}^{N_p }
 {\frac{1}{2}m_i a^2 {\bf{\dot x}}_i^2 }  +
 \frac{G}{a}\sum\limits_{\scriptstyle i = 1 \hfill \atop
  \scriptstyle i \ne j \hfill}^{N_p } {\frac{{m_i m_j }}
  {{\left\| {{\bf{x}}_i  - {\bf{x}}_j } \right\|}}}  + } \right.}  \\
 \left. { + \frac{2}{3}a^2 \pi G\rho _b \sum\limits_{i = 1}^{N_p }
 {m_i {\bf{x}}_i^2 } } \right]dt \\
 \end{array}
\end{equation}

The original interpretation of the conditions (\ref{conditions1}) is
 that at the origin of the Universe the initial positions are free
whereas the velocities are null. Inserting this  in the hierarchical
cosmological context in which the structures grow due to
self-gravity from small deviations  from  homogeneity in an
expanding Universe, the  perturbations grow from different positions
and null velocities. The method does not solve  the equations of
motion, but finds the minimum of the action (\ref{action1}),
partially avoiding the problem of the initial conditions for the
space of velocities (see Sect. \ref{introduction}). The only
observational information that is required are the present
radial velocities and  positions (see Sect. \ref{catalogs}). Owing
to this, the Action Method has  largely been used  to study the
galaxy content of the LG \citep{1989ApJ...344L..53P,
1989ApJ...345..108P, 1990ApJ...362....1P, 1993MNRAS.264..865D,
1995ApJ...449...52P, 2001ApJ...550...87G} or the nearby galaxies
\citep{1994ApJ...429...43P, 1995ApJ...454...15S,
1999ASPC..176..280S, 2001ApJ...552..413G, 2001ApJ...554..104P,
2002ApJ...575....1P, 2002MNRAS.335...53B, 2003MNRAS.346..501B}.

\subsection{Limitations and assumptions}
An intrinsic difficulty of  the variational principle is that it
does not yield  the equations of motion  when  the action
(\ref{action1}) is minimum, but more generally when the derivatives
are null. So a path that minimizes the action (\ref{action1})
satisfies the equations of motion but the opposite is not true. A
path that satisfies the Euler-Lagrange equations could not
correspond to a minimum of the action \citep{1989tadp.book.....W}.
This leads to the problem  of determining the stationary points of
the action integral. The issue  will be addressed in the next
section limited to our specific situation.

Another severe difficulty  of the Action Principle concerns the mass
tracers to use. The methods requires that the Baryonic Matter can be
used as mass tracer and that no lumps of sole Cold Dark Matter are
around altering the estimate derived from baryons. The problem has
long been debated  with controversial results
\citep{1988ApJ...333L..45W, 1983ApJ...267..465D,
2004ApJ...617L..13N, 1995ApJ...443L...1D}.

Despite these points of uncertainty and considering that  we are not
interested in the direct determination of the cosmological
parameters \citep[e.g.][]{1995ApJ...443L...1D}, we  adopt the Action
method. It   allows us  to easily test the results for different
cosmological models.

In the general case of a cosmologically flat Universe, the curvature
of the Universe is given by

\begin{equation}
\label{Auniverse}
    \frac{{\dot a}}{a} = H_0 \left[ {\Omega _0 \left( {1 + z}
     \right)^3  + \Omega _\Lambda  } \right]^{1/2} \,\, ,
\end{equation}

\noindent  where $H_0$ is the Hubble constant, and $\Omega_{0}$ and
$\Omega_{\Lambda}$  the density parameters. The corresponding
orbital equations can be written  as

\begin{equation}
\label{equations1}
    \begin{array}{l}
 a^{1/2} \frac{d}{{da}}\left( {a^{3/2} \frac{{d{\bf{x}}_i }}{{da}}} \right)
  + \frac{{3a^4 \left( {1 - \Omega _0 } \right)}}{{2F_3 }}\frac{{d{\bf{x}}_i }}{{da}} =  \\
  = \frac{{\Omega _0 }}{{2F_3 }}\left[ {\frac{{R_0^3 }}{{M_T }}\sum\limits_{\scriptstyle j =
  1 \hfill \atop
  \scriptstyle j \ne i \hfill}^{N_p } {\frac{{m_j \left( {{\bf{x}}_i  - {\bf{x}}_j }
  \right)}}{{\left\| {{\bf{x}}_i  - {\bf{x}}_j } \right\|^3 }}}  + {\bf{x}}_i } \right] \,\, ,\\
 \end{array}
\end{equation}

\noindent  where $R_0$ is the radius at $t=t_0$ of a sphere
containing the total mass $M_T$ of the galaxies in the solution
assuming an homogeneous mass density. The quantity $F_3$ is derived
from the relation

\begin{equation}
\label{ffunction}
    F_p  \equiv \Omega _0  + a^p \left( {1 - \Omega _0 } \right)
\end{equation}

\noindent with $p=1$ in the case of $\Lambda=1$ and $p=3$ in the
case of $\Lambda\neq0$.  They will be used here only to check the
compatibility with the results derived from the Minimum Action
(\ref{action1}).

\subsection{Analytical representation of the orbits}
A suitable  representation of the orbits to be inserted in  the
system (\ref{equations1}) has been proposed by
\citet{1989ApJ...344L..53P}. They  are expressed as a linear
combination of suitable universal functions of time with unknown
coefficients ${\bf{C}}_{i,n}$

\begin{equation}
\label{sol1}
{\bf{x}}_i \left( a \right) = {\bf{x}}_i \left( {t_0 } \right) +
\sum\limits_{n = 0}^{N - 1} {{\bf{C}}_{i,n} f_n \left( a \right)}
\end{equation}

\noindent for all $i = 1...N_p $, where ${\bf{x}}_i \left( {t_0 }
\right)$ are the present positions of the galaxies, $N_p$  the
number of galaxies or clusters, and $f_n \left( a \right)$ the basis
functions able to satisfy the boundary conditions
(\ref{conditions1}). Classical expressions for the $f_n \left( a
\right)$ are either

\begin{equation}
\label{approxorb1}
    f_n  = a^n \left( {1 - a} \right)
\end{equation}

\noindent that is suggested by \citet{1989ApJ...344L..53P} or

\begin{equation}
    f_n \left( {1 - a} \right)^{s - n} a^n \left( {\frac{{s!}}
    {{n!\left( {s - n} \right)!}}} \right) \wedge n \in \left[ {0,n} \right[
\end{equation}

\noindent that is proposed by \citet{1995ApJ...454...15S}, where $s$
is a parameter and $n<s$.

Another useful expression is the one proposed by
\citet{1993ApJ...411....9G} who assume that  the series representing
the quasi-linear solution has to be as close as possible to the
linear solution and to converge to it in the limit of small
perturbations. Therefore the  leading term of Eqn. (\ref{sol1}) is
the linear approximation of Zel'dovich
\cite[e.g.][]{2002coec.book.....C}. This approximation holds good
only in the earliest stages  of the Universe due to its linear
character. However it has  widely been used to describe density
perturbations over large scales, treating galaxy clusters as
particles. This is possible because the strongly non linear motions
inside a cluster (LG as well as other nearby groups) somewhat
balance each other, so that the motion of the barycentre is
influenced only by its closest neighbourhood. Thanks to it we may
adopt the equations

\begin{equation}
\label{approxorb2}
    {\bf{x}}_i \left( t \right) = {\bf{x}}_{i,0}  + \sum\limits_{n = 1}^N
    {\left( {D\left( t \right) - D_0 } \right)^n } {\bf{C}}_{n,i}
    \,\, ,
\end{equation}

\noindent where the linear-growth function $D$ is taken from
\citet{1977MNRAS.179..351H} and  $D_0$ is

\begin{equation}
    D_0  = D\left( {t_0 } \right) = \left[ {\frac{{\dot a}}{a}\int_0^a
    {\frac{{da}}{{\dot a^3 }}} } \right]_{a_0 } \,\, .
\end{equation}

Finally, another approach has been developed by
\citet{2000MNRAS.313..587N} using the so-called  Fast Action
Minimization Method. In brief,  a Tree-Code algorithm  quickly
computes  the gravitational field contained  in the action
(\ref{action1}) and the solution (\ref{sol1}) is expanded as a
function of the growth factor $D$ \citep{1977MNRAS.179..351H} and
the derivative of the solution (\ref{sol1}). This procedure  is not
used here due to the small number of galaxies we are dealing with
(see Sect. \ref{catalogs}).

As there are many different minima of the action (\ref{action1})
able to reproduce a given set of initial conditions, another version
of the least action method has  recently been proposed by
\citet{2000ApJ...544...21G} and applied to reconstruct the  haloes
of MW and M31 \citep{2001ApJ...550...87G} considered as isolated
systems.

As explained in Sect. \ref{results} we have tried different
algorithms to check the results originating from different minima of
the action.

\section{Neighbourhood of the LG: their dynamical effects}\label{catalogs}

To consider the LG as an isolated system, the forces exerted by the
external galaxies or groups must  be lower than those developed by
the internal galaxies. Typically an external mass $M$ at a
distance $\delta$ will produce a force for unitary mass $
\sim GM/\delta^2 $ if it is sufficiently spherical so that
multi-pole contribution can be neglected. The mutual forces that MW
and M31 exert each other have two main contributions: a mean force
produced by \textsl{distant bodies } (see Table
\ref{tab:ExternalGalaxiesAdopted}) that is  is the target of this
study) and a fluctuating  force due to the dwarf galaxies inside the
LG (Pasetto \& Chiosi 2006 in preparation). To give a rough
idea of the problem, let us consider the LG as made by a number  $N$
of galaxies of mass $M_g$, assumed to be uniformly and spherically
distributed. At any distance $r$ from the centre of the LG
the inner total mass $M_{LG}(r)$ is given by

\begin{equation}
M_{LG}\left( r\right)= NM_g \left( {r\over R_{LG}} \right)^3 \,\, ,
\end{equation}

\noindent where $R_{LG}$ is the total radius of the LG. This
produces a mean force field

\begin{equation}
  <{GM_{LG}\left( r \right) \over r^2 }> \approx {GNM_g\over
  R_{LG}^3}\,r \,\, .
\end{equation}

 \noindent Therefore, the mean
internal force will be greater than the external one  if the
following condition is satisfied

\begin{equation}
{{M_{LG} \over M}} > {R_{LG}^3 \over \delta^2 r } \,\, .
\end{equation}

\noindent Substituting $r$ with $R_{LG}$ we get  $R_{LG}^2
M/\delta^2 M_{LG} < 1$ which, using the entries of Table
\ref{tab:ExternalGalaxiesAdopted}  and the standard values of
$R_{LG}\approx 1.2 Mpc $ and $M_{LG} \approx 2.5 M_\odot$ \citep[see
e.g.][]{1999A&ARv...9..273V, 1999AJ....118..337C}  falls in the
range 0.2 to 0.5. This confirms that the external galaxies do not
dominate the internal dynamics of the LG. However they could have
played an important role either in the past as shown by
\citet{1993MNRAS.264..865D} or even at the present time if one
pushes  the analysis to a deeper level trying to establish the
correlation between torsion and quadrupole moments of the
gravitational field when the assumption of spherical symmetry is
relaxed\footnote{ The contribution  to the force field due to the LG
dwarf galaxies however cannot be estimated in the same way because
the fluctuations are comparable to the instantaneous value.}

\begin{equation}
\label{quadrupolequadrupole} \frac{{d{\bf{I}}_{im} \omega _m
}}{{dt}} = \frac{1}{3}\sum\limits_{j,k}^{} {\varepsilon _{ijk}
q_{jl} Q_{lk} } \,\, ,
\end{equation}
where, $\textbf{I}_{im}$ is the inertia momentum tensor of the LG's
galaxy, $\omega _m$ its frequency of rotation in approximation of
rigid body,
 $q_{jl}$ the quadrupole moment of the selected galaxy,
and $Q_{lk}$ the quadrupole components of the external gravitational
field.
 This is suggested by the planar
distribution we have found and  the results obtained by
\citet{1989MNRAS.240..195R} and \citet{2001ApJ...554..104P}, who
followed the dynamical evolution of the LG galaxies (see Table
\ref{tab:LocalGroupSetOfGalaxies}) considering the influence of
\textsl{external} galaxies and groups. It is worth commenting here
that the concept of LG itself becomes meaningful only when the two
major galaxies are so deeply influencing one another that both the
quadrupole terms and the torsion are of internal origin
\citep{1993MNRAS.264..865D}. Therefore, the evaluation of the
dynamical effects exerted by the dominant galaxies in the LG itself
and  external galaxies and groups is a mandatory,
preliminary step to be undertaken.

The major source of data is the NBG together with many subsequent
implementations. To mention a few, we recall
\citet{1989ApJ...344L..53P},  \citet{1989MNRAS.240..195R},
\citet{1990ApJ...362....1P}, \citet{ 1993MNRAS.264..865D},
\citet{1994ApJ...429...43P}, \citet{1995ApJ...449...52P},
\citet{1999ASPC..176..280S}, \citet{2001ApJ...554..104P}, and
references.

The key data to use here are the distances and the radial
velocities. The uncertainty affecting the distances is particularly
relevant here because it influences very much  the derivation and
prediction of a galaxy orbit and its evolution in turn. Thoughtful
analyses of the importance of good distances in tracking-back  the
orbits of the LG galaxies  are by  \citet{1993AJ....105..886V} and
\citet{1994AJ....107.2055B} to whom the reader should refer for
details.  Their reconstruction  of the orbits of MW and M31 is based
on  the distances of Maffei~1 \citep{1983MNRAS.205..131B} and IC~342
 \citep{1989AJ.....97.1341M} (based in turn on recent determinations
 of the Galaxy extinction). In these studies, the highest
 uncertainty cames from the distance
of Maffei 1. The authors assumed 1.8 Mpc, much shorter than 3.9 Mpc
given by \citet{1988ang..book.....T} and  the recent determination
of 3.5 Mpc  reported by \citet{2003ApJ...587..672F}.

Over the years many efforts have been made to cope with the
uncertainty due to distances. Recently \citet{2002ApJ...575....1P}
showed that a canonical transformation of the action (\ref{action1})
can be used to work directly with the natural boundary conditions at
redshift $z=0$  and the angular position $(l^{\circ},b^{\circ})$ of
a moving observer. Similar attempts have been made either by
changing the orbit functions \citep{1993ApJ...411....9G} or the
action \citep{1998AJ....115.2231S},  or using the method of variable
end-points and the "transversality" condition
\citep{2000ApJ...533...50W}.

In the following we will select a sample of external galaxy groups
for which we will evaluate the dynamical effects on the LG.
Fortunately the uncertainty on the distances for this sample are
found not to significantly affect our results. In any case we will
always keep in mind that distances are the crucial parameter of our
analysis and observational astronomy in general, see e.g.
\citet{2004cgpc.sympE...6C} and \citet{2003LNP...635..265W}.

Given these premises, we briefly review the main properties of the
most massive  galaxies or galaxy groups in both the LG itself and
the surroundings  that could dynamically affect the LG. These are:

\begin{enumerate}
\item \textbf{IC 342 group}: this is the major spiral galaxy of a small group
containing  also NGC 1560 and UGCA 105. Recently
\citet{2002AJ....124..839S} improved upon the distance  using  20
Cepheid stars and derived the distance modulus  $27.58 \pm 0.18$,
i.e. a distance of $3.3 Mpc$. For this group of galaxies we adopt
the total mass of $1.26 \,\, 10^{13} M_ \odot $ and the  radial
velocity of 171 km/s quoted by \citet{1994ApJ...429...43P}.

\item \textbf{Maffei group}: Two galaxies are known under this name, Maffei~1
and Maffei~2. Recent determinations of the distance go from 3 Mpc of
\citet{2003ApJ...587..672F} to 4 Mpc of \citet{1989ApJ...344L..53P}.
We adopt here the  mean value of $3.5 Mpc$ as found in
\citet{1993MNRAS.264..865D}. The total mass of the pair is $6.32
\,\, 10^{12} M_ \odot $. The radial velocity from the MW  is 152
km/s  \citep{1989ApJ...344L..53P}.

\item \textbf{M31 group}: Under this name  we consider the
group containing  M31, M33, IC 10 and  LGS3. The radial velocities
are  taken from \citet{1989ApJ...344L..53P}.  The adopted
total mass mass is $3.16 \,\, 10^{12} M_ \odot $. A smaller
mass for the halo of M31 has been suggested by
\citet{2000ApJ...540L...9E}. It will be taken into consideration
when testing the model results in Sect. \ref{Robustness}.

\item \textbf{M81 group}: This contains  NGC 2366, NGC 2976,
NGC 4236, IC 2574, DDO 53, DDO 82, DDO 165, Holmberg I, Holmberg II,
Holmberg IX, K52, K73, BK3N, Garland, and A0952+69.  Recent
determinations of the kinematics, distances and structures are by
\citet{2002A&A...383..125K}. The total mass is $1.6 \,\,  10^{12} M_
\odot$,  the  mean distance is  3.7 Mpc and the  radial velocity is
130 km/s.

\item \textbf{Cantaurus A (NGC 5128) group}: For the group made of
NGC 5128 and NGC 4945 we adopt the results of
\cite{2005AAS...20717002M}. The distance is 2.7 Mpc, the total mass
is $4.47 \,\, 10^{12} M_ \odot$ and the  radial velocity is  371
km/s.

\item \textbf{Sculptor (NGC 253) group}: This contains  NGC 253, NGC 247,
 NGC 55 and NGC 7793 \citep{1993MNRAS.264..865D}.
The total mass  is  $6.32 \,\, 10^{12} M_ \odot  $, the radial
velocity is  152 km/s. The distance we have adopted is 3.2 Mpc
\citep{1993MNRAS.264..865D}.

\item \textbf{M83 (NGC 5236) group}: The recent study of \citet{2003ApJ...590..256T}
with the ESO VLT yields the distance of 4.5 Mpc and the radial
velocity of 249 km/s. We adopt the total mass of  $7.99 \,\, 10^{11}
M_ \odot $.
\end{enumerate}

 To these groups of galaxies we
have to add of course the MW. The basic data for this sample of
galaxy groups  are given in Table \ref{tab:ExternalGalaxiesAdopted},
which lists the galacto-centric coordinates $l$ and $b$, the
distance, the mass and the radial velocity. The radial
velocities adopted here have been cross-checked with the data in the
catalog by \citet{2005AJ....129..178K}. The differences are taken
into account in the statistical analysis of the results to be
presented in Sect. \ref{Robustness} below. We may anticipate that at
the accuracy level of the present study no significant differences
are found.

\begin{table}
\caption{External   galaxy groups gravitationally influencing the LG
dynamics. }
    \centering
        \begin{tabular}{|c|c|c|c|c|c|}
        \hline
Group &    l   & b &   d & M & $ V_{r}$ \\
        \hline
name   & $^{\circ}$ & $^{\circ}$ & $Mpc$ & $10^{12} M_\odot$ & km/s  \\
        \hline
IC 342    &    138.2   &  10.6   &  3.30 & 12.6  &  171.0 \\
Maffei    &    136.4   &  -0.4   &  3.50 &  6.3  &  152.0 \\
Andromeda &    121.2   & -21.6   &  0.76 &  3.16 & -123.0 \\
Milky Way &      0     &  0      &  0.0  &  2.2  &    0.0 \\
M 81      &    142.1   &  40.9   &  3.7  &  1.6  &  130.0 \\
Cen A     &    309.5   &  19.4   &  2.7  &  4.7  &  371.0 \\
Sculptor  &    105.8   &  85.8   &  3.2  &  6.3  &  229.0 \\
        \hline
        \end{tabular}
\begin{minipage}{0.48 \textwidth}
\footnotesize{The radial velocity is referred to the centre of the
MW. No correction for the motion of the Sun toward  the Local
Standard of Rest is applied because it falls below the accuracy
adopted in this study. The uncertainties on the distances and radial
velocities are omitted.
} \\
\end{minipage}
\label{tab:ExternalGalaxiesAdopted}
\end{table}

These are the eight  dominant galaxy groups. They contains about  90
per cent of all the light and so, if we assume that the light is a
tracer of the mass, as argued in Sect. \ref{minimization}, these are
the key objects whose motion has to be   investigated in order to
derive the force  field acting on the LG.  The analysis will be made
in two steps: (i) first we will consider only the eight objects
listed in Table \ref{tab:ExternalGalaxiesAdopted}; (ii) second,
following \citet{2001ApJ...554..104P}, we will increase the sample
adding a number of galaxies of smaller mass. The new list is shown
in Table \ref{tab:Extendedgalaxyset}.  This would allow us to
estimate the dependence of the results on the number of objects
under consideration or in other words to evaluate how small mass
objects "perturb" the orbits of the large mass ones.

\begin{table}
\caption{Complementary sample of  low mass galaxies.  }
    \centering
        \begin{tabular}{|c|c|c|c|c|c|}
        \hline
Galaxy &     l    & b &    d & $V_r$ &  M   \\
        \hline
N2403    & 149.6  &  29.3  &3.3  & 227  &11.8    \\
N4236    &  12.3  &  47.3  &3.9  & 118  &11.5    \\
Circinus & 311.3  &   3.7  &3.3  & 268  &11.5    \\
N55      & 332.6  & -75.6  &1.6  &  96  &11.5    \\
N7793    &   4.4  & -77.1  &4.1  & 230  &11.5    \\
N300     & 299.2  & -79.4  &2.1  & 101  &11.3    \\
N1569    & 143.6  &  11.2  &2.4  &  47  &11      \\
N404     & 126.9  & -27.1  &3.6  & 115  &10.8    \\
N3109    & 262.1  &  23.0  &1.3  & 194  &10.5    \\
IC5152   & 343.8  & -50.1  &1.7  &  85  &10      \\
SexA,B   & 240.3  &  42.0  &1.4  & 165  &10      \\
N6822    &  25.3  & -18.3  &0.5  &  45  &10      \\
N1311    & 264.2  & -52.6  &3.1  & 425  & 9.8    \\
IC1613   & 129.7  & -60.5  &0.7  &-155  & 9.8    \\
WLM      &  75.8  & -73.6  &0.9  & -65  & 9.7    \\
VIIZw403 & 127.8  &  37.3  &4.5  &  50  & 9.6    \\
SagDIG   &  21.0  & -16.5  &1.1  &  7   & 9.1    \\
PegDIG   &  94.7  & -43.5  &0.8  & -22  & 8.9    \\
LeoA     & 196.8  &  52.4  &0.7  & -18  & 8.8    \\
DDO210   &  34.0  & -31.3  &0.9  & -24  & 8.7    \\
DDO155   & 310.8  &  76.9  &1.5  & 183  & 8.6    \\
Phoenix  & 272.0  & -68.9  &0.4  & -34  & 7.7    \\
        \hline
        \end{tabular}
        \begin{minipage}{0.48 \textwidth}
\footnotesize{ This sample is added to the sample of Table
\ref{tab:ExternalGalaxiesAdopted}. The effect of these galaxies can
be considered as a perturbation. The coordinates $l$ and $b$ are in
$^{\circ}$, the distances $d$ are in Mpc (as in the previous table
we omit the uncertainties on the distances and velocities), the
radial velocity $V_r$ is in km/s, and the galaxy masses are in solar
units. They are expressed as $\log M$.
} \\
\end{minipage}
    \label{tab:Extendedgalaxyset}
\end{table}

\section{Minimizing the action}\label{results}

The main goal of this section is to provide a physical explanation
for the planar distribution of the  galaxies in the LG found in
Sect. \ref{plane}. With the least action principle (Sect.
\ref{minimization}) we want to estimate the influence of the
external galaxy groups (see Table \ref{tab:ExternalGalaxiesAdopted})
on the LG  (see Table \ref{tab:LocalGroupSetOfGalaxies}) and look
for a dynamical justification of the planar distribution we have
found from the mere geometrical analysis. Knowing the
\textsl{external force field} acting on the LG  could indeed cast
light on whether the planar distribution is a temporary situation
holding only at the present time $t=t_0$, whereas the galaxy
phase-space  could tell a different story for all ages different
from $ t_0$.

To this aim, using the galaxies of Table
\ref{tab:ExternalGalaxiesAdopted},    we minimize the action for an
unconstrained  Lagrangian  given by Eqn. (\ref{action1}) \footnote{
A different analysis based on the orbit integrations and proper
motions of  LG galaxies is in preparation (Pasetto \& Chiosi
2006).}.

\begin{figure}
\resizebox{\hsize}{!}{\includegraphics{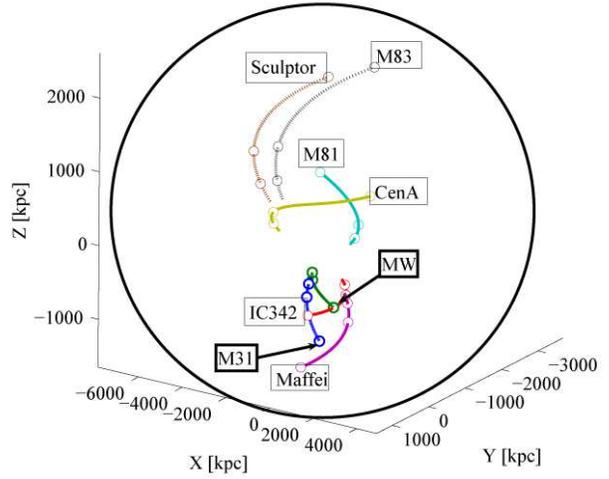}}
\caption{Orbits of the eight dominant galaxies or groups of galaxies
derived from the  minimization of the unconstrained  action of Eqn,
(\ref{action1}).  The orbits are shown as in
\citet{1989ApJ...344L..53P} for the sake of comparison. The orbits
of MW and M31 are highlighted. The circle has a radius $R_0=3.5Mpc$,
see Eqn. \eqref{equations1} and the text for more details. We also
tested the approximation adopted for the analytical representation
of the orbits given by Eqn. (\ref{sol1}) with the equation of motion
of Eqn. (\ref{equations1}). The corresponding orbits are nor shown
as they in practice overlap with the  present ones. Along each orbit
a few selected stages are displayed: the initial stage at $z=60$,
the stage at $z=2$ (the first circle), the stage at $z=1$ (the
second circle) and the stage at the present time  $z=0$ (the third
circle)} \label{fig:GNLfree04}
\end{figure}

We adopt the standard cosmological scenario  with cosmological
constant and the condition $\Omega_0 + \Omega_\Lambda = 1$  we have
already shown in Eqn. (\ref{Auniverse}). The standard cosmological
model is for $H_0 = 72$ km/s/Mpc, $\Omega_0 = 0.3$ and
$\Omega_\Lambda = 0.7$.  The orbits of the eight dominant groups
together with the MW and M31 are followed from redshift
$z=60$ till now. The orbits resulting from the minimization
procedure are presented in Fig.\ref{fig:GNLfree04}.

To avoid solutions with overlapping orbits that cannot  be treated
with the technique  of polynomial expansion of the orbits as a
function of the Universe expansion parameter, we have introduced a
smoothing length $\epsilon$  in the gravitational potential. The
acceleration acting on   the $i^{th}$ particle and measured by a
comoving observer at rest respect to distant matter is given by

\begin{equation}
\label{acceleration}
    {\bf{g}}_i  = \frac{G}{{a^2 }}\sum\limits_{\scriptstyle j = 1 \hfill \atop
  \scriptstyle j \ne i \hfill}^{N_p } {\frac{{m_j \left( {{\bf{x}}_i  -
  {\bf{x}}_j } \right)}}{{\left\| {\left( {{\bf{x}}_i  - {\bf{x}}_j } \right)^2
   + \varepsilon ^2 } \right\|^{3/2} }}}  + \frac{4}{3}a\pi G\rho _b {\bf{x}}_i
\end{equation}

\noindent  where the smoothing length we have adopted is from
\citet{1990ApJ...362....1P}, i.e.  $\epsilon\cong 150$ Kpc.

The smoothing length is the  \textsl{lower limit} below
which the analysis  cannot be pushed, and hence the maximum
 resolution  of the plane thickness. This is dynamically extremely
relevant because it allows us  to avoid un-physical solutions in the
minimization of the action. Indeed the case of  two orbits crossing
each other would inevitably  lead to a merger, which cannot be
properly treated by the technique in use here. Fortunately, it has
recently been proved that  orbit crossing events are less of a
problem as far as the number of solutions able to minimize the
action (\ref{action1}) is concerned \citep{2000ApJ...533...50W}. In
other words, this means that  the dynamics we are going to analyse
is \textsl{collisionless}. It is worth noting that this  description
may not hold at the earliest stages of the Universe (redshift
$z\cong60$) when the galaxy sizes were likely comparable to their
mean separation \citep{1995ApJ...443L...1D}. However, the approach
we have adopted can be  justified in statistical sense: if we
consider our galaxy groups (and dominant galaxies) as mass point,
each with dimension $d \approx \varepsilon $, the smoothing length,
the system of eight point-like galaxy groups can be considered
isolated. This is because the volume occupied by the their orbits,
$ \sim 2000^3 kpc^3 $ (see Fig.\ref{fig:GNLfree04}), is greater than
the total volume occupied by the objects, $ \sim \varepsilon ^3 =
150^3 kpc^3 $. Two objects of radius $\varepsilon $ have an
effective radius of $ 2 \varepsilon $ and a cross section $\sigma  =
4\pi \varepsilon ^2 $. In the very early stages of the Universe,
assuming uniform distribution and random motion for the $N$ objects
of Table \ref{tab:ExternalGalaxiesAdopted}, we can define the mean
free path as $\lambda = \bar v/\bar Vn\sigma$, where the relative
mean velocity of the LG members is nearly equal to the mean velocity
of the group itself, $ \bar V \geq \bar v$. Therefore, the mean free
path of  collisions is $\lambda \approx 1/n\sigma = R^3
/3N\varepsilon ^2 $, about 15 Mpc for the eight massive objects of
Table \ref{tab:ExternalGalaxiesAdopted}, i.e. larger than the
dimensions we are interested in. This means that we expect
collisions to rarely affect the motion of the objects in the sample
\citep[but see e.g.][]{2005Natur.433..389D, 2005nfcd.conf..211D}.
The same considerations cannot be extended to the total sample given
by Tables \ref{tab:LocalGroupSetOfGalaxies},
\ref{tab:ExternalGalaxiesAdopted} and  \ref{tab:Extendedgalaxyset}
because  of the very large range of  masses spanned by the objects.
It is worth recalling here that the constraints given by Eqn.
(\ref{conditions1}) refer to the present time, whereas back in the
past  ($z=60$) the positions were random and the velocities null. In
other words, but for the very initial stages, as soon as the
collisionless description can be applied, the results we have
presented are a good description of the real galaxy motions.

Finally, the motions can be referred to the barycentre of
the system to simply and correctly follow the Universe expansion.
The new positions and velocities are given in Table
\ref{tab:initalcondition1} (and in Fig.\ref{fig:GNLfree04}). For
instance, referred to the barycentre of the group ${\bf x}_G =
[1117,\, 986, \, 975]$ kpc, the  velocity vector of M31 at the
present time is  $ {\bf{v}}_{M31} \left( {t_0 } \right) = [\rm
125.5,\,  - 25.9,\,  - 85.0]$ km/s and the projection of the
velocity vector onto the line joining the barycentre to the M31
position is $ {\bf{v}}_{rad,M31}^{\left( G \right)} \left( {t_0 }
\right) =  - 15.4 $ km/s. This is one of the many possible velocity
vectors matching this radial projection, however,  it is the only
one \textsl{minimizing the action } of Eqn. (\ref{action1}).

\begin{table*}
\caption{Velocities and positions of the galaxies used in the
minimization procedure.}
    \centering
        \begin{tabular}{|c|c|c|c|c|c|c|}
        \hline
Galaxy &$v_x^G$&$v_y^G$&$v_x^G$&$x^G$&$y^G$&$z^G$\\
        \hline
name   & km/s & km/s & km/s & kpc & kpc & kpc  \\
        \hline
IC 342     &-294.7 &  80.0 &-21.1 & 1309.6 &  1176.0 & -368.2 \\
Maffei     &-359.1 &  97.1 &-66.1 & 1426.0 &  1427.6 & -999.7 \\
Andromeda  & 125.5 & -25.9 &-85.1 & -742.5 &  -381.6 &-1255.0 \\
Milky Way  & 117.3 & -67.0 &-65.7 &-1108.5 &  -986.0 & -975.3 \\
M 81       &-391.3 &  49.8 & 97.4 & 1098.3 &   731.9 & 1447.3 \\
Cen A      & 725.4 &-200.7 & 83.4 &-2728.4 & -2951.1 &  -78.4 \\
Sculptor   & 822.1 & -51.7 &149.5 &-1044.7 &  -760.5 & 2216.2 \\
M 83       & 618.0 &-250.6 & 89.1 &-3807.0 & -3684.5 & 1409.4 \\
        \hline
        \end{tabular}
                \begin{minipage}{0.6 \textwidth}
\footnotesize{The masses have been taken from Table
\ref{tab:ExternalGalaxiesAdopted}. The positions are referred to the
barycentre  of the sample located at ${\bf x}_G  = \left\{ {{\rm
1117}{\rm , 986}{\rm , 975}} \right\}$ kpc.
} \\
\end{minipage}
    \label{tab:initalcondition1}
\end{table*}

This kind of analysis based on the same sample of Table
\ref{tab:ExternalGalaxiesAdopted} has already been made by
\citet{1993MNRAS.264..865D} to derive  the proper motion of M31.

At this stage,   an interesting question can be posed. Is
there any correlation between the plane found in Sect. \ref{plane}
and the action minimization? In other words how would the action
minimum change if the plane is added as constraint to the whole
procedure?

Using the  representation of the plane given by Eqn.
(\ref{vectors2}), we impose that the positions of M31 and MW are
those given by the observations, the same for the radial velocity of
M31, and that the present-day velocity vectors of the two galaxies
belong to the plane. From a dynamical point of view, the latter
hypothesis is not equivalent to that adopted by
\citet{2005PASJ...57..429S}. They assumed a central \&
plane motion for the LG galaxies and derived the dynamical history
of the LG. Here \textsl{no assumption} is made \textsl{neither on
the angular momentum or torsion} of the LG, nor the
quadrupole-quadrupole interaction, nor the inertia tensor of our
main galaxies (see Eqn. \eqref{quadrupolequadrupole} and
\citet{1993MNRAS.264..865D} for similar considerations). Finally,
owing to the expected minimum thickness of the theoretical plane
caused by the smoothing length,  we allow for the possibility that
the velocity vector cannot to exactly belong  to the  plane,
although very close to it. From a mathematical point of view, this
means that a anholonomic constraint has to be chosen instead of the
pure holonomic one used in Sect. \ref{plane}. Therefore the
holonomic constraint evaluated \textsl{only at the time} $t=t_0$ and
expressed by

\begin{equation}
\label{condition2}
v_x n_x  + v_y n_y  + v_z n_z  \cong 0
\end{equation}

\noindent is replaced by the more suitable condition

\begin{equation}\label{condition3}
\left\| {a\left( {\frac{{dx}}{{da}}n_x  + \frac{{dy}}{{da}}n_y +
\frac{{dz}}{{da}}n_z } \right) + n_x x + n_y y + n_z z} \right\| \le
\varepsilon \,\, ,
\end{equation}

\noindent where $\varepsilon  \to 0$. This is not a genuine
anholonomic constraint because it has to be satisfied only at the
extremal points of all possible trajectories and not along the whole
path, i.e.

\begin{eqnarray}
\label{conditions4}
    \left\{ \begin{array}{l}
 \delta {\bf{x}}_i \left( {t = 0} \right) \ne 0 \wedge
 \mathop {\lim }\limits_{a \to 0} a^2 {\bf{\dot x}}_i  = 0 \\
 \mathop {\lim }\limits_{t \to t_0 } \delta {\bf{x}}_i  = 0 \wedge
 {\bf{\hat n}} \cdot {\bf{x}}_i \left( {t = t_0 } \right) = 0 \,\, .\\
 \end{array} \right.
\end{eqnarray}

\noindent This assumption allows us to avoid more complicated
minimization procedures, e.g. \citet{SalCrom} for the Constrained
Minimum Action Principle.

\begin{figure*}
\resizebox{\hsize}{!}{\includegraphics{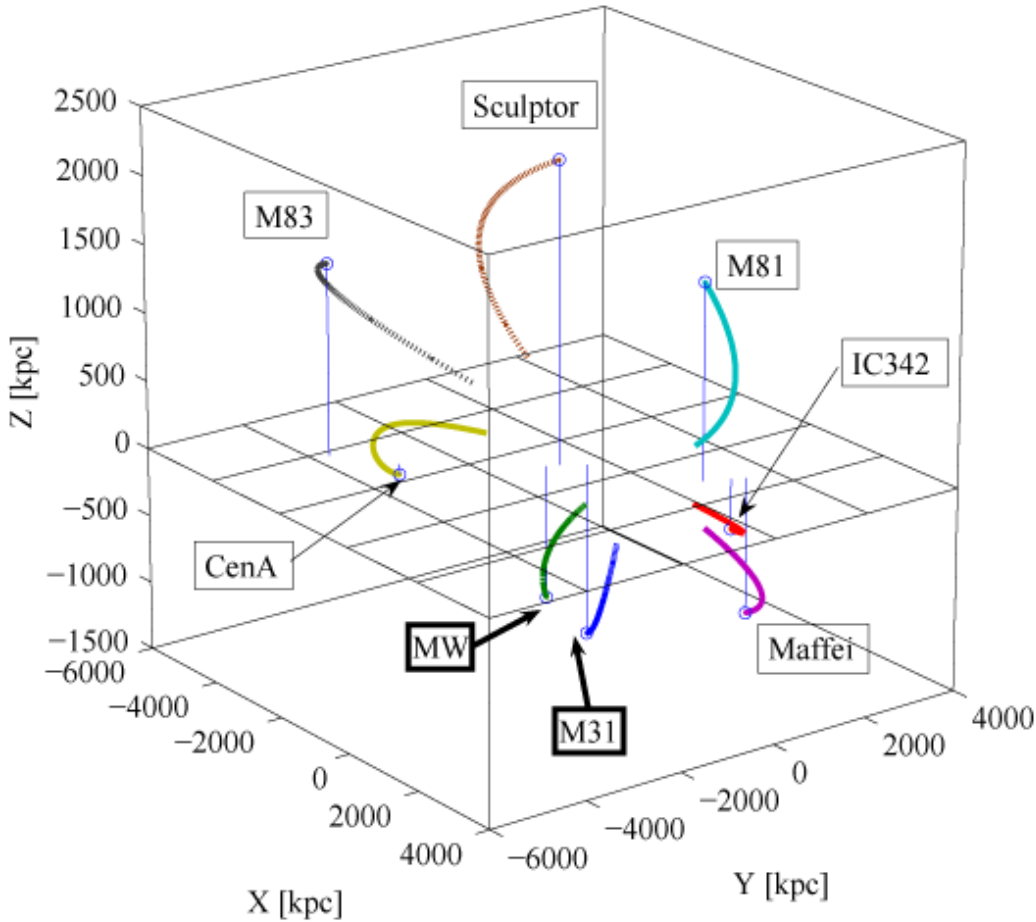}}
\caption{Orbits of the galaxy groups and the two dominant galaxies
derived from the constrained minimization of the action. The
constraints are given by relations (\ref{condition2}) and
(\ref{condition3}). The dots and circles along each orbit have the
same meaning as  in Fig.\ref{fig:GNLfree04}. The circles correspond
to the present age.
 } \label{fig:GNLcondizionato04}
\end{figure*}

The solution for the minimization  is shown in
Fig.\ref{fig:GNLcondizionato04} and the companion Table
\ref{tab:initalcondition2}. In general the  velocities remain the
same as before but for M31 and MW  for which they significantly
change. This simply reflects the  constraint that  at $t=t_0$ the
two galaxies in the phase space   belong to the manifold expressed
by the conditions (\ref{conditions1}) and  (\ref{condition2}).
\textsl{The most interesting and important result is that
now the action  is significantly lower than in the previous case. We
suspect that profound implications are hidden in this result. We may
provisionally conclude that the orbits of MW and M31 lie on a plane
and that this likely coincides with that derived from the
geometrical analysis.}

Before proceeding further, there are a number of considerations to
be made here:

\begin{itemize}

\item As both the Lagrangian of   the action integral
(\ref{action1}) and the set of galaxies in use (Table
\ref{tab:ExternalGalaxiesAdopted}) are  let unchanged,   the space
of possible  constraints of the first minimization  must contain
also the one of the second minimization. Therefore, in the first
case the procedure has not been able to locate the absolute minimum
because of the large dimensions of the minimization space (roughly
120 dimensions). The minimum  found in the first case can  be
considered either as a local minimum or a saddle. The saddle
situation is a classical problem with this kind of minimization and
often masks the true solution \citep[see e.g.][ and the case of  NGC
6822 orbit]{2003MNRAS.346..501B, 1994ApJ...429...43P}. This somewhat
proves that the second minimization is more correct than the first
one.

\item The attempts to get the motion of M31 based on this methods
cannot give physically sounded results  without an adequate estimate
of the errors. Therefore, the proper motion derived by
\citet{1993MNRAS.264..865D} and \citet{2005PASJ...57..429S} has to
be considered as one of the possible values. We will touch upon this
point in more detail testing the stability of our results (see
below).

\item Velocity vectors belonging to the plane of MW, M31
and all other dwarf galaxies of the LG correspond  to a solution
more realistic than the one given by the un-constrained
minimization. This  does not, however, ensure that the solution is
\textsl{stable}. Even assuming that in the LG the  force field is
dominated by  the binary interaction between MW and M31, this does
not automatically exclude that the two galaxies are simply
undergoing a fly-by encounter and that they will not depart from
each other in the future. The  existence of a \textsl{dynamically
justified plane } must be proved  by the analysis of the external
forces acting on the LG at all ages.
\end{itemize}

In the section below we will examine in detail the robustness of the
result  just obtained and  the implications for the force field
acting on the galaxies of the LG.

\begin{table}
\caption{Velocities derived from the Constrained Minimum Action. }
    \centering
        \begin{tabular}{|c|c|c|c|}
        \hline
Galaxy &$v_x^G$&$v_y^G$&$v_z^G$\\
        \hline
name   &km/s & km/s & km/s \\
        \hline
IC 342      &-297.6 &  83.2 &-24.5 \\
Maffei      &-352.2 &  94.7 &-61.8 \\
Andromeda   & -22.8 & -26.0 &  2.7 \\
MW          &   7.1 &  55.7 &-64.5 \\
M 81        &-387.1 &  48.8 & 97.0 \\
Cen A       & 722.4 &-215.1 & 79.2 \\
Sculptor    & 827.1 & -50.1 &142.9 \\
M 83        & 622.1 &-245.4 & 88.7 \\
        \hline
        \end{tabular}
        \begin{minipage}{0.48 \textwidth}
\footnotesize{The positions are taken from literature and
are left unchanged. See the text for more details.
} \\
\end{minipage}
    \label{tab:initalcondition2}
\end{table}

\section{Robustness of the solution}\label{Robustness}
We proceed now to test the solution we have obtained  as far as
three points of weakness are concerned.

\begin{enumerate}

\item \textbf{The minimization method.} How much the result depends on the
minimization method in use? To answer this question first we adopt
three different minimization algorithms, second we  change  the
random number generator and repeat the analysis several times
because in the so-called optimization problems in multidimensional
spaces one deals with statistical results that have to be tested
against the adopted statistical method (see below).\\

\item \textbf{Orbit representation.} The analytical representation of
the orbits  used in Eqns. (\ref{action1}) and (\ref{equations1}) to
obtain the final velocities is a critical aspect of the whole
problem. Like in other studies,  \citep[see for
instance][]{1989ApJ...344L..53P,1993MNRAS.264..865D}, the orbit
traced-back by the action (\ref{action1}) agree with those derived
with the equation of motion (\ref{equations1}). However, the tangent
to the orbits at $a=1$, i.e. $t=t_0$,  strongly depends on the
number of terms considered in the series  expansion of Eqn.
(\ref{approxorb1}). To be physical meaningful, the result must be
independent of analytical expressions in use. Therefore, we will
check whether the results remain the same using  Eqn.
(\ref{approxorb2}).\\

\item \textbf{Number of galaxies in the sample}. Finally, it is worth
testing whether the   results vary with the number of galaxies that
are taken into account. To this aim  we re-determine  the velocities
for the extended set of galaxies which includes also all the objects
listed in Table \ref{tab:Extendedgalaxyset}.
\end{enumerate}

\subsection{Minimization methods}
In this subsection we test the possible dependence of the results
from the minimization method. In literature there are several
algorithms  of  optimization  in multidimensional spaces. In
general, a great advantage is possible  if the gradient in the
action (\ref{action1}) can be calculated

\begin{equation}
\label{gradient}
    \nabla _{{\bf{C}}_{i,n} } S = m_i \int\limits_0^{t_0 }
    {\left[ { - \frac{d}{{dt}}\left( {a^2 {\bf{\dot x}}_i } \right)
    + a{\bf{g}}_i } \right]f_n dt} \,\, ,
\end{equation}

\noindent  where the gravitational acceleration is given by Eqn.
(\ref{acceleration}). This possibility makes the methods based on
the nonlinear conjugate gradient \citep{1986nras.book.....P} very
efficient to derive an unconstrained optimization. Alternatively,
for a small sample of galaxies,  one can invert the matrix of the
second derivatives as described in \citet{1995ApJ...449...52P}. This
was indeed the first method to be adopted.  However, there are other
methods that are suited to the aims  and are also widely used in
literature. In the following we apply some of them to our problem.

\begin{itemize}

\item \textbf{Differential evolution}. This is a widely used genetic algorithm
\citep{storn95differential,fleiner97parallel}. It stands on a series
of guess vectors, ${\bf{x}}_1 ,...,{\bf{x}}_n $, of real numbers
(\textsl{genes}). At each iteration, each ${\bf{x}}_i $ randomly
selects  the integers $\left\{a,b,c\right\}$ to build up the mate
${\bf{y}}_i  = {\bf{x}}_a  + \gamma \left( {{\bf{x}}_b - {\bf{x}}_c
} \right)$ where $\gamma  = {1 \mathord{\left/ {\vphantom {1 2}}
\right. \kern-\nulldelimiterspace} 2}$ is a scaling factor. Then
${\bf{x}}_i $ is linked to ${\bf{y}}_c$ according to the value of
the crossing probability (constantly equal to ${1 \mathord{\left/
{\vphantom {1 2}} \right. \kern-\nulldelimiterspace} 2}$) producing
the ${\bf{z}}_c$. Finally ${\bf{x}}_i $ competes against ${\bf{z}}_i
$ for the position of ${\bf{x}}_i $ in the population. This method
is customarily adopted to address problems with integer values.
However  it has been tested to produce  reasonable values even in
our case. The method is somewhat time consuming.

\item \textbf{Down Hill Simplex}. This method has been developed
by \citet{Nelder:1965:SMF} and it has become the subject of many
text books \citep{1986nras.book.....P}. The action (\ref{action1})
is evaluated for some values or vertices  of the coefficients
${\bf{C}}_{n,i} $ of the solutions (\ref{sol1}) and minimized. Once
the vertex coefficients have been determined, the method starts
from the highest vertex, ${\bf{x}}_h  $, and folds itself
about the centroid, ${\bf{x}}_c $, of the remaining points to a new
point, ${\bf{x}}_a $, in such a way that $ {{\left\| {{\bf{x}}_a  -
{\bf{x}}_c } \right\|} \mathord{\left/
 {\vphantom {{\left\| {{\bf{x}}_a  - {\bf{x}}_c } \right\|} {\left\|
 {{\bf{x}}_c  - {\bf{x}}_h } \right\|}}} \right.
 \kern-\nulldelimiterspace} {\left\| {{\bf{x}}_c  - {\bf{x}}_h } \right\|}} $
 is equals  to 1. If
${{\bf{x}}_a } $ falls in  between the highest and lowest vertex, it
replaces ${{\bf{x}}_h } $ and the process goes on. If ${{\bf{x}}_a }
$ is the highest vertex, the method constructs ${{\bf{x}}_b } $ so
that $ {{\left\| {{\bf{x}}_c  - {\bf{x}}_b } \right\|}
\mathord{\left/
 {\vphantom {{\left\| {{\bf{x}}_c  - {\bf{x}}_b }
 \right\|} {\left\| {{\bf{x}}_c  - {\bf{x}}_h } \right\|}}} \right.
 \kern-\nulldelimiterspace} {\left\| {{\bf{x}}_c  - {\bf{x}}_h } \right\|}} $
 is equals to $1/2$. If ${{\bf{x}}_b } $ is lower than
${{\bf{x}}_h } $, the method replaces ${{\bf{x}}_h } $ with
${{\bf{x}}_b } $ and goes on, otherwise all vertex of the simplex
are moved toward the lowest vertex ${{\bf{x}}_l } $ producing a new
vertex ${\bf{x'}}_i$ so that $ {{\left\| {{\bf{x}}_i  - {\bf{x'}}_i
} \right\|} \mathord{\left/
 {\vphantom {{\left\| {{\bf{x}}_i  - {\bf{x'}}_i } \right\|}
 {\left\| {{\bf{x}}_i  - {\bf{x}}_l } \right\|}}} \right.
 \kern-\nulldelimiterspace} {\left\| {{\bf{x}}_i  - {\bf{x}}_l } \right\|}} $ gets equal
  to $1/2$.

\item
\textbf{Simulated Annealing}. This is a random-walk search method.
It generates a set of random starting points and for each one it
takes a random direction along which  to move on. If the new point
yields a better solution it is accepted, otherwise the method
calculates the probability $P$ and compares it to a random value
$n\in\left[0,1\right]$.  If $n<P$,  the point is accepted even if
the solution is not improved. The probability is given by $ P = \exp
\left[ {b\left( {i,\Delta f,f_0 } \right)} \right]$ where $b$ is
defined by $ b\left( {i,\Delta f,f_0 } \right) = {{ - \Delta f\log
\left( {i + 1} \right)} \mathord{\left/
 {\vphantom {{ - \Delta f\log \left( {i + 1} \right)} {10}}} \right.
 \kern-\nulldelimiterspace} {10}}$, with $i$ the iteration number, ${\Delta f}$ the
change in the target function, and $f_0$ thee value of the target
function in  the previous iteration \citep{1986nras.book.....P}.
\end{itemize}

We have applied all the three  methods using  different random seeds
and suitably changing the parameters of the method in use. This has
required a large amount of computer time, however  without leading
to significant changes in the solutions  for the velocity vectors of
our galaxies. Indeed all  velocity vectors  differ from
each other only by few km/s. The same  for the action minimum.

Finally as we were more interested in the physical meaning
of the result, we preferred \textit{not} to look for the absolute
minimum for the action. Instead,  following \citet{Efron:1986:BMS}
we considered the \textsl{mean minimum value } of the action
resulting from a statistical  interpretation of the results. As far
as the orbits are concerned we performed many numerical experiments
in which a fraction of galaxies of our sample has been randomly
dropped from  the configuration space and their distances randomly
varied within the uncertainties, and finally the minimization has
been recomputed. The final adopted values are for the statistical
interpretation of the orbits. The results are presented in Table
\ref{tab:valorimediati}. A few simulations have been also calculated
using the largest set of galaxies of Table
\ref{tab:Extendedgalaxyset}. All results are fully compatible with
those of Table \ref{tab:valorimediati}.  To fully and properly
perform a statistical analysis one should also vary the velocities
within their uncertainties. The estimated time cost of such analysis
is very large. Fortunately, the small mass for most of galaxies in
Table \ref{tab:Extendedgalaxyset} ensures us that the final result
would not significantly differ from the the present one.

\subsection{Analytical representation of the orbits}
 As in the real space the tangent to the solution for the orbits
given by Eqn. (\ref{sol1}) is expected to strongly depend on the
number of terms in   the polynomial expansion of the solution, we
have extended it up to the ninth term ($N=9$)

\begin{equation}
\label{sol2} {\bf{x}}_i \left( a \right) = {\bf{x}}_i \left( {t_0 }
\right) + \sum\limits_{n = 0}^{8} {{\bf{C}}_{i,n} f_n \left( a
\right)} \,\, .
\end{equation}

\noindent In  the same way we have extended the analytical
representation of Eqn. (\ref{approxorb2}) up to  $N=4$ and $N=8$

\begin{equation}
\label{sol3}
    {\bf{x}}_i \left( t \right) = {\bf{x}}_{i,0}  + \sum\limits_{n = 1}^8
    {\left( {D\left( t \right) - D_0 } \right)^n } {\bf{C}}_{n,i}
    \,\, .
\end{equation}

To show the largest difference caused by the order of the analytical
expansion of  Eqn. (\ref{sol3}),  in Table
\ref{tab:parametrizzazione2} we list the velocity vectors. As
expected, the difference is rather small and comparable to that of
Table \ref{tab:valorimediati}.

Finally, we would like to remark   that  the analytical
representation of the orbits and the minimization procedure  do not
affect the plane position which is only revealed  by the Principal
Direction analysis. Small errors in the plane determination could
result from the  approximation made in  Eqn. \eqref{condition3}.
Therefore, they are only due to the velocity section of the tangent
bundle of the configuration space where our smooth Lagrangian
function \eqref{action1} is defined.

\begin{table}
    \caption{Mean value of the velocities derived from the
    analysis explained in the text.                                                                                                                                                                                                                     }
    \centering
        \begin{tabular}{|l|r|r|r|}
        \hline
Galaxy &$v_x^G\pm \sigma_x$&$v_y^G\pm \sigma_y$&$v_z^G\pm \sigma_z$\\
        \hline
name   &km/s & km/s & km/s \\
        \hline
IC 342    &$-298 \pm 10 $&  $81 \pm 6$ & $-22 \pm 8$ \\
Maffei    & $-353 \pm 15$ & $ 99 \pm 11$ & $-65 \pm 15$\\
Andromeda & $-22 \pm 17$ & $-26 \pm 14$ &  $2 \pm 18$ \\
MW        & $  9 \pm 9  $&$  57 \pm 13$ &$-65 \pm 14 $\\
M 81      &$-380 \pm 16$ &$  49 \pm 7 $ &$ 96 \pm 11$ \\
Cen A     &$ 722 \pm 22$ &$-210 \pm 19 $&$ 72 \pm 24 $\\
Sculptor  &$ 820 \pm 12 $&$ -49 \pm 18 $&$ 148 \pm 31$\\
M 83      & $620 \pm 14$ &$-245 \pm 16$ & $87 \pm 6$ \\
        \hline
        \end{tabular}
        \begin{minipage}{0.48 \textwidth}
\footnotesize{The errors are \textsl{relative quantities}
    due to the statistical approach we have adopted.} \\
\end{minipage}
    \label{tab:valorimediati}
\end{table}

\begin{table}
\caption{Local minimum velocity vectors for the solutions given by
Eqn. (\ref{sol3}). }
    \centering
        \begin{tabular}{|c|c|c|c|}
        \hline
Galaxy &$v_x^G$&$v_y^G$&$v_z^G$\\
        \hline
name   &km/s  & km/s & km/s \\
        \hline
IC 342      &-295.807  &  79.9568 &-25.0462 \\
Maffei      &-360.432  &  97.0    &-67.9    \\
Andromeda   & -22.7    & -25.9    &  2.6    \\
MW          &   7.2    &  57.2    &-66.3    \\
M 81        &-392.5    &  49.7    & 98.4    \\
Cen A       & 727.9    &-200.6    & -5.3    \\
Sculptor    & 824.5    & -51.7    &150.6    \\
M 83        & 620.4    &-250.5    & 95.8    \\
\hline
\end{tabular}
\begin{minipage}{0.5 \textwidth}
\footnotesize{These values are fully compatible with those listed in
Table \ref{tab:valorimediati}.
} \\
\end{minipage}
\label{tab:parametrizzazione2}
\end{table}

\subsection{Richness of the galaxy sample}
\label{setofgalaxies} 
\begin{figure}
\resizebox{\hsize}{!}{\includegraphics{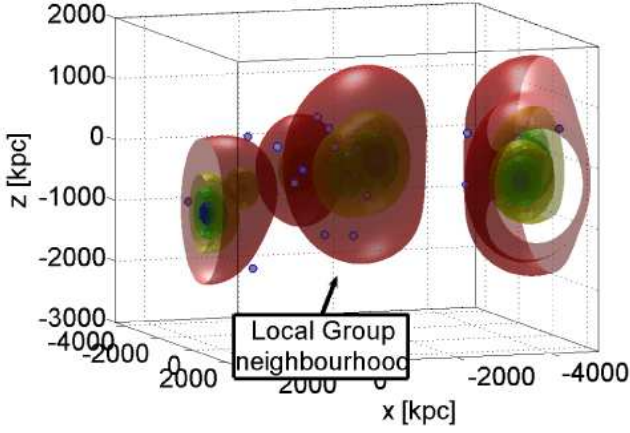}}
\caption{Surfaces of constant gravitational potential for the
galaxies and groups of galaxies given in  Tables
\ref{tab:Extendedgalaxyset} and \ref{tab:ExternalGalaxiesAdopted}.
The colour code makes it evident how the spherical symmetry
dominates. The position of the LG is indicated.
 } \label{fig:Isocaps01}
\end{figure}

Finally, we evaluate the effect  of changing
the number of galaxies used  to determine the force field acting on
the LG. The analytical representation of the orbits is the one of
Eqn. (\ref{sol1}) and the total galaxy sample is the sum of Tables
\ref{tab:ExternalGalaxiesAdopted} and \ref{tab:Extendedgalaxyset}.
The new  velocity vectors are found to be very similar to those of
Table \ref{tab:valorimediati}. This check is very time consuming and
more subjected to the \textsl{errors in the distance values} of
Table \ref{tab:Extendedgalaxyset}.  Recalling that the number of
interactions scales as $N_p^2$, the  computational time is greatly
increased. The little change in the velocity vectors that is found
with the enlarged sample of galaxies is particularly interesting
because it implies that the forces acting on the LG galaxies are
primarily of internal origin. This is  best illustrated by the shape
of the equal gravitational potential surfaces and the field lines
around each galaxy or galaxy group of the large sample (Tables
\ref{tab:ExternalGalaxiesAdopted} and \ref{tab:Extendedgalaxyset})
that are shown in Fig.\ref{fig:Isocaps01}. The gravitational
potential around each object is nearly spherical and the field lines
are nearly radial (see also Fig.\ref{fig:GNLcondizionatoVF09}
below). With these external gravitational fields,  no significant
effect can be produced  on the LG galaxies even if the distance
errors in some of the galaxies of \ref{tab:Extendedgalaxyset} could
be relevant.

\section{The external force field}\label{EF}
\begin{figure}
\resizebox{\hsize}{!}{\includegraphics{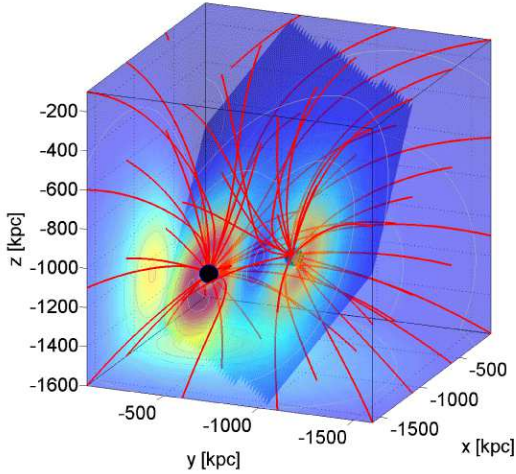}} \caption{The
force field showing the radial behaviour of the streamlines of the
MW and M31 projected on the  geometrical  plane that minimizes the
distances of the LG galaxies.} \label{fig:GNLcondizionatoVF09}
\end{figure}

In this section we examine in more detail the spatial shape of the
gravitational potential (and associate field lines)  shown in
Fig.\ref{fig:Isocaps01} in order to better understand the effect  of
the external force field on the  planar distribution of the LG
galaxies discovered in Sect. \ref{plane}. To this aim we limit
ourselves to consider only the subgroup strictly belonging to the LG
neighbourhood (see Table \ref{tab:ExternalGalaxiesAdopted}).  As
pointed out by \citet{1993MNRAS.264..865D} it is plausible that in
the past
 dynamics and angular momentum (that ultimately  depend on the
quadrupole moment of the external gravitational potential, see Eqn.
\eqref{quadrupolequadrupole}) of the LG have been strongly
influenced not only by the two most massive spiral galaxies, M31 and
MW, but also by  the nearest external galaxies. \textsl{At the
present time} the dominant feature of the force field is the
two-body interaction between the two dominant spirals. The
present-day situation is illustrated  in
Fig.\ref{fig:GNLcondizionatoVF09} where we see the typical dipolar
structure of the field line generated by MW and M31. This ensures
that only the central forces are governing the dynamics of the two
major galaxies.

The plane corresponding to the central motion of MW and M31 not
necessarily  has to coincide  with  the plane derived from
geometrical arguments. Indeed the dynamics of the  dwarf galaxies
satellites of MW and M31 could be fully incompatible with the planar
distribution indicated by the geometry. The latter indeed could be a
transient  situation with no correlation with  the orbital plane
revealed by the minimum action. In Fig.\ref{fig:GNLcondizionatoVF09}
we plot the dipole force field at the present time ($t=t_0$)  over
the geometrical plane. The dipole force field is the one derived
from  the minimum action (Sect. \ref{results}) whereas the
geometrical plane is the one derived from Eqn. (\ref{vectors2}). On
the geometrical plane the present-day dipole structure of the force
field  is still evident.

To cast light on the dynamical significance of the geometrical
plane, instead of investigating   the dynamics due to the central
force between the two major galaxies, we must clarify  the influence
of the external galaxies on both the geometrical plane and the
central force plane.

 Following \citet[][]{1989MNRAS.240..195R}, the force field minimizing
 the action can be expressed as

\begin{equation}
{\bf{F}}\left( {\bf{r}} \right) = G\sum\limits_{\scriptstyle j = 1 \hfill \atop
  \scriptstyle j \ne \{ MW,M31\}  \hfill}^{N_p }
  {\frac{{m_j \left( {{\bf{r}} - {\bf{r}}_j } \right)}}{{\left( {\left\|
  {{\bf{r}} - {\bf{r}}_j } \right\|^2  + \varepsilon ^2 } \right)^{3/2} }}}
\end{equation}

\begin{figure}
\resizebox{\hsize}{!}{\includegraphics{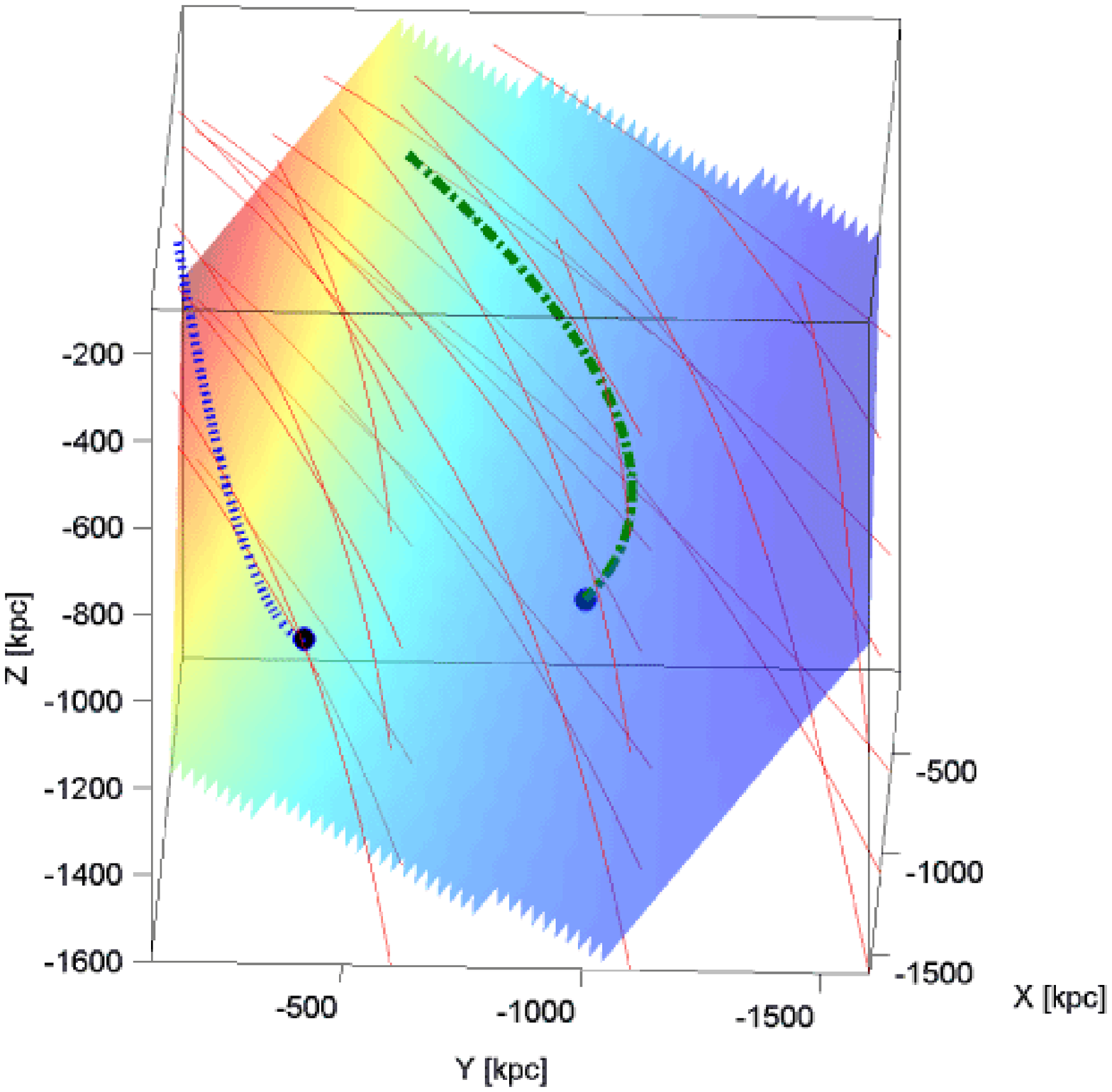} }
\resizebox{\hsize}{!}{\includegraphics{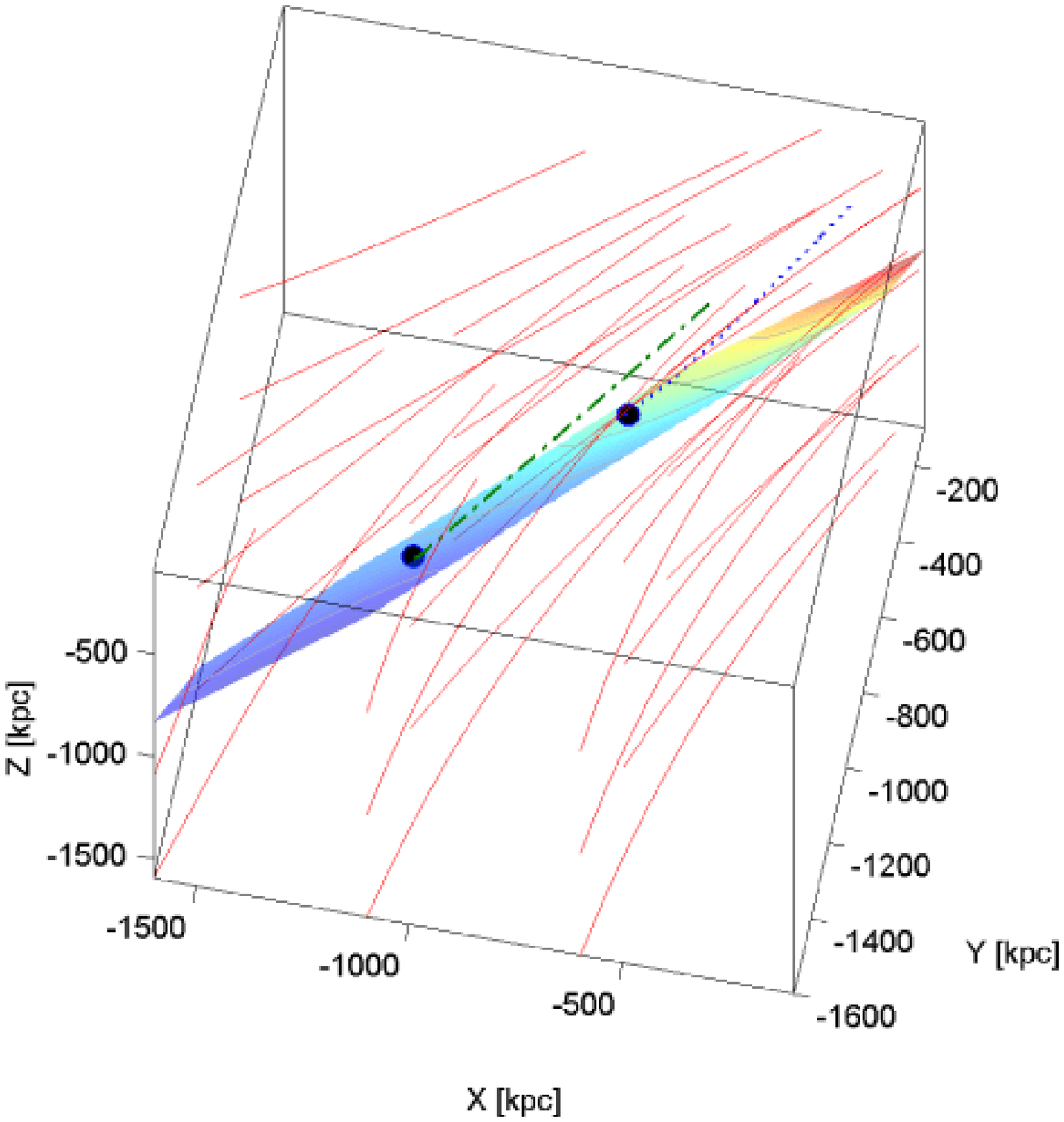} }
\caption{{\it Top Panel}: Face-on view of the plane minimizing the
action and the geometrical distances of the LG galaxies together
with the field line of the gravitational interaction exerted by the
external galaxies on the LG.  {\it Bottom Panel}: The same but
viewed edge-on. Note how the field lines are nearly parallel to the
plane. See the text for more details}
\label{fig:GNLcondizionatoVF21_VF22}
\end{figure}

\noindent Starting from this, we seek to understand how the external
force field has influenced the past and present dynamics of the LG.
This is the only way we can check the  stability of the planar
distribution  by tightening together the static hint from the
geometry and the dynamical hint from the orbits. The
surprising result  is shown in
Fig.\ref{fig:GNLcondizionatoVF21_VF22}, which displays the face-on
(top) and edge-on (bottom) projection of the geometrical plane, the
lines of the force field and the position of the MW and M31.

The \textsl{force field of the external galaxies that minimize the
action is clearly parallel to the plane. This means that the force
perturbing the central dynamical forces  of the LG is dragging the
plane toward the direction orthogonal to the  normal direction, in
other words the force field is pulling the plane along it}. The
geometrical and central force planes coincide within  the
uncertainty of the procedure we  adopted to bring all this into
evidence. As we have not made any special  assumption for the
angular  momentum of the LG this finding is very interesting.
\textsl{Indeed the solution with velocity vectors belonging to the
geometrical plane does not give any hint about the
inclination  of the central force plane}. An infinite set
of different possibilities is compatible with the assumption that
the velocity vectors of M31 and MW belong to the geometrical plane,
but the solution that minimize the action show that the
orbits should be parallel to the geometrical plane, at least from
$z=2$. Interestingly enough, the coincidence between the geometrical
and central force planes stems from the minimum action. The external
field might have been significantly different at epochs earlier than
$z=2$ \citep{1993MNRAS.264..865D} but ever since it did not change
too much. The orbits derived from the minimum action (Sect.
\ref{results}) belong  to the geometrical plane. The two planes have
been coincident for a large fraction of the Hubble time (nearly
all).

This makes it possible to constrain the proper motions of the dwarf
galaxies external to the LG \citep{2005PASJ...57..429S}, or more
realistically to constrain the dwarf galaxies (mostly irregulars) of
the LG that are preferentially located  at large distances from the
two major galaxies and are subjected to a lower degree of spherical
symmetry in the force field due to the dark matter halo of the two
massive spirals, see
 \citep{2003A&A...405..931p}. The force field lines shown
 in Fig.\ref{fig:GNLcondizionatoVF21_VF22}  are likely to describe,
 with no  significant variations with time,   the  direction and
intensity of the force since very early epochs, say $z\cong2$.

\section{Conclusions}\label{concl}
In this paper we have investigated the spatial distribution
of the galaxies in the LG. They seem to share a common plane. The
plane has been first geometrically singled out and then dynamically
interpreted.

The geometrical discovery has been made using the Principal
Component Analysis for an un-weighed sample. The plane we find does
not coincide with those recently suggested by
\citet{2005PASJ...57..429S}, on the base of the sky-projected
distribution, and \citet{1979ApJ...228..718K} and \citet{
1999IAUS..192..447G} from dynamical studies limited to the  MW and
M31 sub-systems.

Major uncertainty of our analysis is the issue of the mass
tracers. Indeed  the luminous objects have been used to derive from
merely geometrical arguments the plane minimizing their distances to
this.
 We have reasonably supposed that pure Dark Matter haloes do not
likely exist in the LG at least for redshifts $z\leq 2$. However,
the presence of such pure Dark Matter haloes, while not affecting
the determination of the geometrical plane, could  affect  on the
orbital motions and the planar distribution of galaxies that we have
derived from the dynamical analysis.

The second part of the article is devoted to investigate the
dynamics of the two dominant galaxies of the LG, MW and M31, under
the action of external forces exerted by nearby galaxies or galaxy
groups external to the LG. The aim is to prove whether the planar
distribution we have found is stable in time or it is a mere
coincidence that will disappear in the future. We have demonstrated
that the force field acting on the dynamics of the LG as a whole is
compatible with the geometrical plane, in other words with a planar
distribution that remain stable for very long periods of time. The
result is that there is no external force field able to destroy the
planar distribution. Once again the possible presence of undetected
pure Dark Matter haloes may be a drawback of the present study.

Putting together the information from the geometrical and
dynamical analysis the real thickness of the plane is estimated to
be about 200 kpc.

\begin{figure}
\resizebox{\hsize}{!}{\includegraphics{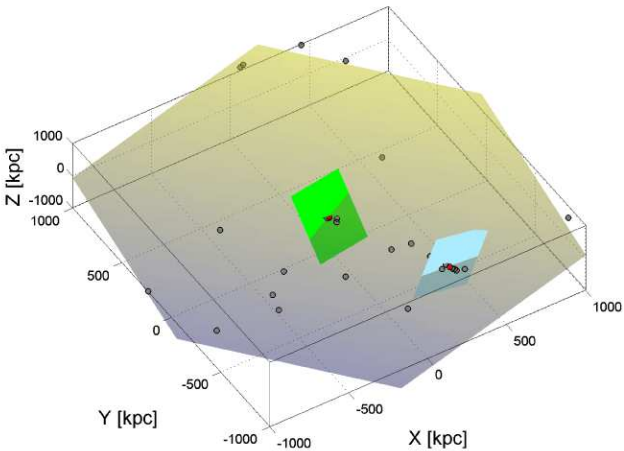}}
\caption{The planes for the dwarf galaxies around the MW
\citep{2005A&A...431..517K} and  M31 \citep{2006AJ....131.1405K}
compared to our plane. Unfortunately, we cannot show the plane of
\citet{2005PASJ...57..429S} as no equation for it has been given explicitly by
the authors.}
 \label{fig:PlanesComparison}
\end{figure}

Furthermore, despite the uncertainties, there is a strong
hint that the motion of the galaxies inside the LG is due to a
central force field.  The long membership of the two spiral galaxies
to the plane, for about 10 Gyr, provides some information on the
virialization state of the LG and the proper motions. However,
despite the extensive use  by \citet{2005PASJ...57..429S}, the
presence of a plane should be adopted only to determine of proper
motions  of the dwarf galaxies in the outskirts of the LG,  as close
to the dominant spirals the gravitational potential    and the
fly-by encounters are far from being collisionless and
scatter phenomena can complicate the description (see Pasetto \&
Chiosi 2006, in preparation). We would like to remark that from the
results of Table \ref{tab:valorimediati} we can derive the proper
motions expected for M31. Following \citet{2006A&A...451..125V}
(their Eqn. (4)) and assuming that the Sun is at rest in the Local
Standard of Rest system, the velocity system $(U,V,W)$ of Eqn. (4)
in \citet{2006A&A...451..125V} is aligned with our velocity system.
Therefore  the peculiar velocity of M31 is simply the velocity given
in Table \ref{tab:valorimediati} once translated into the velocity
system centered on the MW. We obtain $\left( {\mu _l ,\mu _b }
\right) = \left( {-3.03 \cdot 10^{-4}, -3.92 \cdot 10^{-4} } \right)
{\rm arcsec \, yr^{ - 1} } $, fully compatible with
\citet{1993MNRAS.264..865D}.
 As expected the proper motions turn out to be small due to the
nearly parallel motion of M31 with respect to the MW.  This result
also indirectly confirms that the numerical errors  in our analysis
are quite small as we derive a radial velocity of M31 equal to $-
115 \,{\rm km \, s^{-1}} $  which is virtually identical to the one
we have initially  imposed, i.e. the observational value $v_r = -
113 \,{\rm km \, s^{-1}} $.

Finally in Fig.\ref{fig:PlanesComparison} we plot and
compare with ours  the planes found   by \citet{2005A&A...431..517K}
around the MW and by \citet{2006AJ....131.1405K} around M31.  First
of all, the planes found by \citet{2005A&A...431..517K} and
\citet{2006AJ....131.1405K} differ from each other (have different
normals).  While the normal to the plane around MW seems to be
aligned with the direction  toward   M31 this is not the case for
the plane around M31 whose normal does not align with the direction
 towards the MW. Second, both planes do not coincide with our plane. The
problem is rather complicated and in part mirrors the different
methods that have been adopted in the three studies to address the
subject and to determine the planes. They stem indeed from different
dynamical considerations. In brief, the gravitational interaction
among galaxies can be studied in two regimes. On long distances it
can be investigated as a collisionless process best described by a
statistical method. This is the approach of the present study as
well as of many others in literature. Basing on this,  we have
derived the spatial structure and evolution of the MW and M31
complexes together with their satellites.  On short distances, the
gravitational interaction is best modelled by collisional dynamics.
This is the typical case of studies on the orbits of satellites of
big mother galaxies and the effect of these latter on the spatial
distribution of their nearby dwarf galaxies. In this context
\citet{2006AJ....131.1405K} suggested for the M31 group a tidal
break up of a  pre-existing dominant  galaxy. This fact could
partially explain the lack of collinearity between the planes for
the M31 and MW complexes. However, it is likely that also the
progenitor of M31 has been influenced by the external force field we
have analysed in the present study. Furthermore, the seemingly
parallel pull of the external field minimizing the action could have
influenced, roughly in the same manner, both the proto-M31 and the
proto-MW (if any). As a final consideration on this issue, it may
well be that on a local scale, i.e. when studying the dynamical
behaviour of the satellite galaxies belonging to the MW and M31
complexes, they seem to lie on planes with different inclination,
but the generalisation of this result to the whole LG may not be
correct. On the grand scale of the whole LG all galaxies seem to
belong to a rather thin slab with its own (different) inclination
which is dot destroyed by external actions.  A study of the
LG including  both the collisional  effects among interacting dwarf
galaxies and the tidal effects by the external field on the orbits
of the dwarf galaxies is underway (Pasetto \& Chiosi 2006, in
preparation).

 In this paper we tried to set the ground for future analysis
in which the two regimes for the gravitational interaction can be
taken into account. Specifically, we tried to impose constraints to
the tangent bundle space of a smooth collisionless Lagrangian
function generated by a group of galaxy clusters, using equations
that stem  from  a local analysis and the probable existence of a
plane revealed by merely geometrical considerations. The Lagrangian
is constrained only at the time-boundaries of its region by means of
the velocity space over there. In other words, the action
minimization is constrained at the two temporal extremes: at the
origin of the Universe by means of the Hamilton Principle (in the
Peebles interpretation) and at the present time by means of the
observational velocities and positions.  Therefore, the action
minimization here is one of the first order, i.e. \textit{a free,
unconstrained minimization but for the temporal extremes with no
other constraint along the minimization path}. The next step forward
would be to include the collisional regime. Unfortunately, in this
case the dynamical interactions easily loose memory of the initial
conditions. Therefore the dynamics of the LG must be investigated by
imposing \textit{additional new and suitable constraints on the
tangent bundle space of the Lagrangian}. Precious hints come from
the stellar content of the mother and dwarf galaxies of the LG
\citet{2006AJ....131.1405K}. In other words, the dynamical history
of individual galaxies is recorded not only in their space-time
geodesics or their past orbits (that anyhow are still poorly
determined despite the many efforts in literature) but also in their
past history of star formation and chemical enrichment on which we
know lots more, see e.g. the reconstruction of this information made
over the past two decades deciphering the Colour-Magnitude Diagrams
of their stellar populations \citep[see][ for a recent review of the
subject and references therein]{2005ARA&A..43..387G}. The target is
ambitious but feasible.

\begin{acknowledgements}
This study was financed  by the University of Padua by means of a
post-doc fellowship to S. Pasetto and the Department of Astronomy by
providing  indirect logistic support (computers, infrastructure,
etc). No financial resource has been allocated to the project by the
Italian Ministry of Education, University  and Research (MIUR).
\end{acknowledgements}

\appendix

\section{}
Here we describe  the simple method used to minimize the
\textsl{distance} expressed by Eqn. (\ref{distancef}) subjected to
the constraint given in Eqn. (\ref{constr1}) and the constraint that
the  plane we are looking for must  be orthogonal the vector joining
M31 and MW, i.e. ${\bf{a}} = \left( {a_x ,a_y ,a_z } \right)$. This
additional holonomic constraint can be written as ${\bf{n}} \cdot
{\bf{a}} = 0$. The problem can be further simplified assuming as
origin of the coordinate system of reference one of the two major
galaxies.  This means that the constant $c$ of Eqn. (\ref{distance})
can be set to zero $c=0$. Therefore ${\bf{n}} \cdot {\bf{a}} = 0$
and $\left| {\bf{n}} \right|^2 - 1 = 0$ are the system of
independent constraints. The distance defined by Eqn.
(\ref{distancef}) in the Euclidean space $E^3$ will present a
stationary point ${\mathbf{n}}$ if the gradients of the constraints
in ${\mathbf{n}}$ are able to express $\nabla D$ as a linear
combination of them, i.e are a basis for $N_{\mathbf{n}} E^3 $, the
space orthogonal to the manifold of ${\mathbf{n}}$. This leads to

\begin{equation}
2\sum\limits_{g = 1}^{N_p } {\left\langle {{\mathbf{n}}{\mathbf{,x}}^{\left( g \right)} }
 \right\rangle {\mathbf{x}}^{\left( g \right)} }  =
 2\lambda _1 {\mathbf{n}} + \lambda _2 {\mathbf{a}} \,\, .
\end{equation}

\noindent For simplicity we express the position as
\begin{equation}
{\bf{A}}\left( { = A_{ij} } \right) \equiv 2\sum\limits_{g = 1}^N {\left( {x_i x_j }
 \right)^{\left( g \right)} }
\end{equation}

\noindent and shortly rewrite  the previous equation as
\begin{equation}
\label{App1} \left\langle {{\bf{A}},{\bf{n}}} \right\rangle  =
\lambda _1 {\bf{n}} + \lambda _2 {\bf{a}} \,\, .
\end{equation}

\noindent This  is a non homogeneous system of linear equations with
parameters $\lambda _1 $ and $\lambda _2 $. We can also write

\begin{equation}
 \left\langle {{\bf{A}},{\bf{n}}} \right\rangle  - \lambda _1
{\bf{n}} = \lambda _2 {\bf{a}}
\end{equation}

\noindent and re-cast the system as

\begin{equation}
\left( {{\bf{A}} - \lambda _1 {\bf{I}}} \right){\bf{n}} = \lambda _2
{\bf{a}}
\end{equation}

\noindent The rank of the symmetric matrix in $\Re^3$ is 3, so  that
$\det \left[ {{\bf{A}} - \lambda _1 {\bf{I}}} \right] \ne 0$ which
means that $\lambda _1 $ is \textsl{not} an eigen-value for
$\bf{A}$. In this case a solution of the system \ref{App1} surely
exists because the rank of the matrix ${\bf{M}} \equiv \left(
{{\bf{A}} - \lambda _1 {\bf{I}}} \right)$ is the same as the rank of
the same matrix completed with the column $\left\{\lambda _2
{\bf{a}}\right\}$: $\left\{ {{\bf{M}},\lambda _2 {\bf{a}}}
\right\}$.  This  can be translated into a  condition on the
parameter $\lambda _1 $, not on $\lambda _2 $. The solutions form a
vectorial space and they can be determined  with the Kramer's
method.

However, as we are more interested in the components of the
ortho-normal vector ${\bf{n}}$ more than in the Lagrange's
multipliers $\lambda _1 $ and $\lambda _2 $, we can work directly on
the system (\ref{App1}).  Taking the scalar product of both members
of (\ref{App1})  we obtain

\begin{equation}
\begin{array}{l}
 \left\langle {\left\langle {{\bf{A}},{\bf{n}}} \right\rangle ,{\bf{n}}} \right\rangle  =
 \left\langle {\lambda _1 {\bf{n}},{\bf{n}}} \right\rangle  +
 \left\langle {\lambda _2 {\bf{a}},{\bf{n}}} \right\rangle  \\
 \left\langle {{\bf{An}},{\bf{n}}} \right\rangle  = \lambda _1 \left\langle {{\bf{n}},{\bf{n}}} \right\rangle  + {\bf{n}}\left\langle {{\bf{a}},{\bf{n}}} \right\rangle  \\
 \left\langle {{\bf{An}},{\bf{n}}} \right\rangle  = \lambda _1  \,\, , \\
 \end{array}
\end{equation}

\noindent where once again we find  the dependence of the solution
on $\lambda _1 $.  We  can avoid all this by explicitly  writing the
components as

\begin{equation}
\begin{array}{l}
 n_j \left[ {\sum\limits_{g = 1}^N {\left( {x_j^g \sum\limits_{i = 1}^3 {n_i x_i^g } } \right)} }
  \right] = \lambda _1 \;j = 1,2,3 \Leftrightarrow  \\
 \left\{ \begin{array}{l}
 n_1 \left[ {\sum\limits_{g = 1}^N {\left( {x_1^g \sum\limits_{i = 1}^3 {n_i x_i^g } } \right)} }
  \right] = \lambda _1  \\
 n_2 \left[ {\sum\limits_{g = 1}^N {\left( {x_2^g \sum\limits_{i = 1}^3 {n_i x_i^g } } \right)} }
  \right] = \lambda _1  \\
 n_3 \left[ {\sum\limits_{g = 1}^N {\left( {x_3^g \sum\limits_{i = 1}^3 {n_i x_i^g } } \right)} }
  \right] = \lambda _1   \,\, , \\
 \end{array} \right.  \\
 \end{array}
\end{equation}

\noindent which means that
\[
\left\{ \begin{array}{l}
 n_1 \left[ {\sum\limits_{g = 1}^N {\left( {x_1^g \sum\limits_{i = 1}^3 {n_i x_i^g } }
  \right)} } \right] = n_2 \left[ {\sum\limits_{g = 1}^N
  {\left( {x_2^g \sum\limits_{i = 1}^3 {n_i x_i^g } } \right)} } \right] \\
 n_1 \left[ {\sum\limits_{g = 1}^N {\left( {x_1^g \sum\limits_{i = 1}^3 {n_i x_i^g } }
  \right)} } \right] = n_3 \left[ {\sum\limits_{g = 1}^N
   {\left( {x_3^g \sum\limits_{i = 1}^3 {n_i x_i^g } } \right)} } \right] \\
 n_2 \left[ {\sum\limits_{g = 1}^N {\left( {x_2^g \sum\limits_{i = 1}^3 {n_i x_i^g } }
  \right)} } \right] = n_3 \left[ {\sum\limits_{g = 1}^N
   {\left( {x_3^g \sum\limits_{i = 1}^3 {n_i x_i^g } } \right)} } \right]  \, \, .\\
 \end{array} \right.
\]

\noindent This is a system whose solutions $\left\{ {n_1 ,n_2 ,n_3 }
\right\}$ yield   the vector we are  looking for. The numerical
solution  is straightforward.

\bibliographystyle{apj}
\bibliography{LocalGroup}
\end{document}